\RequirePackage{fix-cm}

\documentclass{svjour3}       
\smartqed  
\usepackage{graphicx}
\usepackage{color}
\definecolor{mygreen}{rgb}{0.0,0.6,0.45}

\usepackage{amsmath}
\usepackage{amsfonts}

\usepackage{hyperref}

\usepackage[font=small,labelfont=bf]{caption}

\newcommand{\Tr}{\operatorname{Tr}}

\newcommand{\beq}{\begin{equation}}
\newcommand{\eeq}{\end{equation}}

\begin{document}

\title{Large deviations at level 2.5 for Markovian open quantum systems: quantum jumps and quantum state diffusion}
\titlerunning{Level 2.5 large deviations for open quantum systems} 

\author{Federico Carollo$^{1,\dagger}$ \and Juan P. Garrahan$^{2,3}$ \and Robert L. Jack$^{4,5}$}
\authorrunning{Federico Carollo, Juan P. Garrahan, Robert L. Jack} 

\institute{%
$^\dagger$ Corresponding Author: federico.carollo@itp.uni-tuebingen.de\\
		$^1$	Institut f\"{u}r Theoretische Physik, Universit\"{a}t T\"{u}bingen, Auf der Morgenstelle 14, 72076 T\"{u}bingen, Germany \\
	   $^2$	School of Physics and Astronomy, University of Nottingham, Nottingham, NG7 2RD, UK \\
           $^3$  Centre for the Mathematics and Theoretical Physics of Quantum Non-Equilibrium Systems,
            University of Nottingham, Nottingham, NG7 2RD, UK	   \\
           $^4$	Department of Applied Mathematics and Theoretical Physics, University of Cambridge, Wilberforce Road, Cambridge CB3 0WA, UK\\
	   $^5$	Yusuf Hamied Department of Chemistry, University of Cambridge, Lensfield Road, Cambridge CB2 1EW, UK\\
}

%

\maketitle

\begin{abstract}
We consider quantum stochastic processes and discuss a level 2.5 large deviation formalism providing an explicit and complete characterisation of fluctuations of time-averaged quantities, in the large-time limit.
We analyse two classes of quantum stochastic dynamics, within this framework. The first class consists of the quantum jump trajectories related to photon detection; the second is quantum state diffusion related to homodyne detection. 
For both processes, we present the level 2.5 functional starting from the corresponding quantum stochastic Schr\"odinger equation and we discuss connections of these functionals to optimal control theory.
\end{abstract}

\section{Introduction}

The time-evolution of closed quantum systems is unitary, deterministic, and governed by  Schr\"odinger equations. 
By contrast, open quantum systems (see e.g.~\cite{Breuer2002,Gardiner2004} for reviews) are constantly interacting with their environment.  In such cases, dynamics is no longer unitary due to dissipation and mixing effects, and to the flow of information into the (often infinitely many) degrees of freedom of the environment. Within Markovian and weak coupling approximations, such system dynamics are implemented by Lindblad (or Lindblad-Gorini-Kossakowski-Sudarshan) dynamical generators \cite{Lindblad1976,Gorini1976},  which describe evolution under the assumption that the system-bath interaction is not monitored in any way. The resulting quantum dynamics is deterministic and probability conserving but in general  non-unitary. 

However, modern experiments can monitor correlations between the system dynamics and the environment through suitable measurement processes \cite{PhysRevLett.57.1696,PhysRevLett.57.1699,PhysRevLett.106.110502,Gleyzes:2007aa,PhysRevLett.56.2797}.
For example, a single experiment can yield a time-record of observations, from which the behaviour of the system and the bath can be fully reconstructed. This time-record of events is stochastic because of the fundamental laws of quantum mechanics. It is associated with a quantum trajectory \cite{Belavkin1990,Dalibard1992,Gardiner1992,Carmichael1993} that  specifies the evolution of the system state conditioned on the given time-record. Averaging this state over all possible time-records (trajectories) recovers the dynamics generated by a Lindbladian. Going beyond the average, information about dynamical fluctuations is available by analysing stochastic quantum trajectories.

In this paper, we explain how to characterise large dynamical fluctuations of quantum stochastic processes by means of the theory of large deviations (LD) \cite{Derrida1998,Lebowitz1999,denHollander,Giardina2006,Lecomte2007b,Garrahan2007,Lecomte2007,Touchette2009,Jack2010,Chetrite2015,Chetrite2015b,jack2019ergodicity}. In particular, we present results about very general LD functionals which encode information about fluctuations of measurement outcomes.  This includes general linear and non-linear functions of the  quantum state of the system. 
We address the two main classes of measurement processes that monitor the interaction of a quantum system with its own environment. One class involves the detection of bath quanta emitted by the system -- such as photons or particles -- and gives rise to discontinuous quantum jump trajectories \cite{Dalibard1992,PhysRevA.35.198,Plenio1998,Breuer2002,Gardiner2004}. The other involves the continuous monitoring of homodyne currents associated to bath operators, which gives rise to quantum state diffusion  \cite{Belavkin1990,Plenio1998,Breuer2002,Gardiner2004}. We note that similar equations arise also when describing weak or strong measurements of system observables, see for example \cite{PhysRevLett.116.110401}.

The functionals that we derive and  discuss represent the counterpart of level 2.5 LD functionals in jump and diffusion processes \cite{Maes2008,Barato2015,Chetrite2015b,Hoppenau2016,Bertini2018} for quantum stochastic processes. A short presentation of the results for quantum jump processes has appeared before in \cite{carollo2019}. We now present (in Sec.~\ref{sec:outline}) an overview of our main results, including (in Sec.~\ref{subsec:structure}) an outline of the structure of  the following Sections.

\section{Outline}
\label{sec:outline}

\subsection{Scope}
\label{outline:scope}

We consider Markovian open quantum dynamics in which the state of the system is described by a reduced (system) density matrix $\rho(t)$ that is obtained by tracing out the environment.  We focus on finite-dimensional quantum systems described by means of a Hilbert space $\mathbb{C}^n$, where $n$ is the maximum number of orthogonal (basis) states of the space. The quantum state $\rho$ is then a Hermitian $n\times n$ matrix with non-negative eigenvalues and $\Tr \rho = 1$.
In the Markovian limit, the dynamics of $\rho$ is given by the Lindblad (or Lindblad-Gorini-Kossakowski-Sudarshan) equation \cite{Breuer2002,Gardiner2004,Lindblad1976,Gorini1976}
\beq
\dot\rho = {\cal L}(\rho) 
\label{outline:lindblad-eq}
\eeq
with
\beq
{\cal L}(\rho) = -i [H,\rho] + \sum_{i=1}^M\left( L_i \rho L_i^\dag - \frac12 \left\{ \rho,L^\dag_i L_i \right\} \right)\, ,
\label{outline:lindblad-op}
\eeq
where $H$ is the system Hamiltonian and the $L_i$ are jump operators that depend on the coupling to the environment. 
This assumption is sufficiently general to cover the dynamics of several interesting quantum systems in contact with their environment \cite{Gardiner2004}. We sometimes refer to the generator ${\cal L}$ as the {\em Lindbladian}. 

\begin{figure}[t]
\includegraphics[width=11cm]{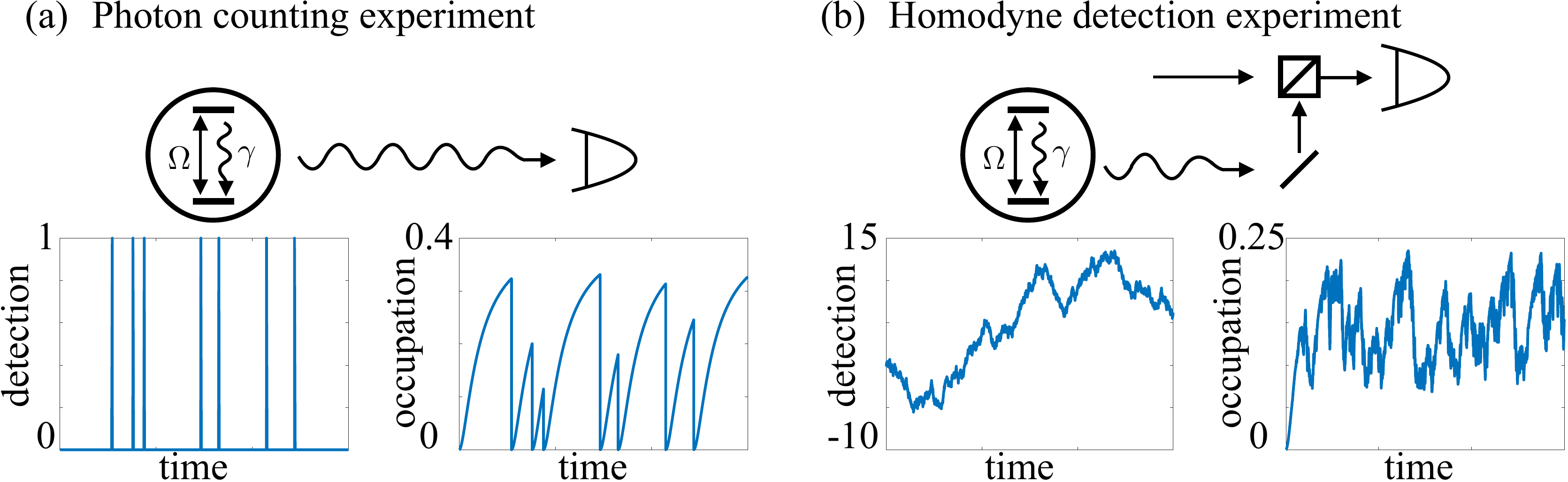}
\caption{Sketch of two different experiments for open quantum systems. We show illustrative results for the simple two level system. (a) Photon counting experiment: a detector reveals the emission (or the absence of  emissions) of a photon. The time-record of this measure is a sequence of times at which photons are detected. For each such event detected, the state of the system changes abruptly and collapses to its de-excited state, as it can be seen in the plot of the occupation of the excited state. (b) Homodyne detection experiment: the output light field emitted by the atom interferes with a local oscillator field and is measured by the detector. The detection outcome consists of the time-integrated homodyne current, measuring the intensity of the quadrature of the light field. The quantum state, in this setting, does not undergo sudden jumps but is instead diffusing as measured by the average occupation.  }
\label{Fig1}
\end{figure}

From the density matrix, it is possible to compute all observable properties of the system.  In this work we go further, by considering correlations between system observables and measurements that are made in the environment, as well as time-correlations in the stochastic system dynamics. Two general settings are considered: (i) correlations between system properties and the statistics of quantum jumps, corresponding to emission/absorption (for example of photons) into/from the environment, see Fig.~\ref{Fig1}(a); and (ii) correlations between system properties with the measurement of homodyne currents see Fig.~\ref{Fig1}(b).  The theories for these two cases are different in their details, although there are common features.  The case of quantum jump detection is discussed in Section~\ref{sec:jump} while homodyne measurements are discussed in Section~\ref{sec:diff}.

\subsection{Unravelling}

Some correlations between system and environment can be analysed by tilted variants of Eq.~\eqref{outline:lindblad-eq}, see \cite{Esposito2009,Garrahan2010,Budini2010,Hickey2012,Chetrite2012,Znidaric2014,Znidaric2014b,carollo2018,cilluffo2020microscopic}.  Here we take a different approach, which is to unravel the joint dynamics of the system and the environmental measurements.  This enables access to a larger set of dynamical observables and correlations. The theory is based on the stochastic evolution of a pure-state density matrix $\psi_t$, which is a Hermitian $n\times n$ matrix for which one eigenvalue is $+1$ and the others are all zero. This matrix evolves by a stochastic process \cite{Belavkin1990,Gardiner2004} which we write (schematically) as
\beq
\mathrm{d}\psi_t = b(\psi_t) \mathrm{d}t + \mathrm{d}\omega_t\, ,
\label{outline:sde}
\eeq
where $\mathrm{d}\omega_t$ represents a random (stochastic) increment for $\psi$, see below for details. One could also consider the stochastic dynamics of a mixed (non-pure) density matrix. This would be necessary, for instance, in those cases in which the measurement performed on the quantum system can have degenerate outcomes.  In these cases, we expect the general theory to be the same, however, some details, such as the space of states in which the stochastic process takes place, would need to be modified (see also Section \ref{sec:notation}).
Averaging over the noise with suitable initial conditions for $\psi$, a general result is that
\beq
\mathbb{E}[\psi_t] = \rho(t)\, ,
\eeq
where $\mathbb{E}$ is an expectation value for the stochastic process.
Hence the (quantum-mechanical) average of any system observable $A$ can be obtained as $\langle A(t) \rangle = \mathbb{E}[ \Tr(A\psi_t)]$.

Note that this construction makes use of a density matrix $\psi$ that remains normalised at all times $\Tr(\psi_t)=1$.  
Other descriptions of the unravelled dynamics may be expressed in terms of states (or matrices) whose norm (or trace) is also time-dependent~\cite{Gardiner2004}.  In what follows, we will construct probability distributions for $\psi_t$, for which it is convenient that this object remains normalised.

Every trajectory of the stochastic process \eqref{outline:sde} is associated with a time-record for the environmental measurements.  For example, if the measurements involve photon counting then the noise $\omega_t$ causes jumps in $\psi_t$, and the number of these jumps is, for example, the number of emitted photons.  Writing ${N}_t$ for the number of jumps between time 0 and time $t$, 
this allows computation of observables such as
\beq
\langle N_t A(\tau) \rangle = \mathbb{E}[ N_t \Tr(A\psi_\tau) ] \, .
\eeq
This is an example of an observable quantity that depends on correlations between the system observable $A$ and the environmental measurement $N_t$, see for example Ref.~\cite{PhysRevLett.116.110401}.

The unravelled system also allows access to objects that are not immediately experimentally observable.  In particular, quantum mechanical expectation values are linear functions of $\psi$ but one may also consider objects that are non-linear.  For example, in bipartite systems (decomposed into subsystems $A$ and $B$), the entanglement of the (pure) density matrix $\psi$ is
\beq
S_E(\psi) = -\Tr_{\rm A}( \chi(\psi) \log\chi(\psi) )\, ,
\eeq
where $\Tr_{\rm A}$ denotes a partial trace of subsystem A and $\chi(\psi) = \Tr_{\rm B}(\psi)$.  Then
\beq
{\cal S}_E(t) = \mathbb{E}\left[ S_E(\psi_t) \right]
\eeq
measures the average value of the entanglement shared by the two subsystems.   This quantity, obtained as an average over time records, is nowadays receiving a lot of attention \cite{Nahum2017,Nahum2018,PhysRevB.98.205136,Keyserlingk2018,PhysRevX.9.031009,PhysRevB.101.104301,ippoliti2020entanglement,alberton2020trajectory,nahum2021measurement}.

\subsection{Large deviations at level 2.5}

Our focus in this work is on large deviations of time-integrated quantities.  A simple example would be $N_t$, the number of emitted photons, as above.  For large $t$, the distribution of $N_t$ is sharply-peaked, in the sense that its mean is proportional to $t$, while its standard deviation is proportional to $\sqrt{t}$.  Large deviation theory \cite{Giardina2006,Lecomte2007b,Garrahan2007,Lecomte2007,Touchette2009,Jack2010,Chetrite2015,Chetrite2015b,jack2019ergodicity} can be used to analyse the {\em rare events} where $N_t$ differs significantly from its mean value, as $t\to\infty$.
The statistical properties of these events are described by large deviation theory at {\em level 1}, within the classification of Donsker and Varadhan \cite{DonVarI,DonVarII,DonVarIII,DonVarIV}.

Here we are concerned with large deviations at a more abstract level of theory, which is called in the LD jargon {\em level 2.5}.  To explain this, we first consider {\em level 2}, which motivates us to define the empirical measure for the (pure-state) density matrix $\psi$.  This is 
\beq
\mu_t(\psi) = \frac1t \int_0^t \mathrm{d}t' \, \delta( \psi - \psi_{t'} ) 
\label{outline:mu}
\eeq
where we have introduced a Dirac delta function in the space of density matrices, see later sections for details.  
For a trajectory of (\ref{outline:sde}), this $\mu_t(\psi)$ measures (roughly speaking) the fraction of the time interval $[0,t]$ that the system spent in state $\psi$.    

We assume throughout that the system has a unique stationary state, hence for large times $\mu_t(\psi)$ should converge to some $P_\infty(\psi)$, which is the steady state distribution for $\psi$.  Large deviation theory at level 2 allows computation of the probability that $\mu_t$ differs significantly from $P_\infty$.  However, this level 2 theory is not sufficient for our purposes, for example it cannot capture the probability distribution of quantities like $N_t$.  The solution to this problem is to consider the joint statistics of $\mu_t$ and the empirical fluxes $Q_t$, which correspond to time-averaged jump rates for all possible quantum jumps.  The precise definition of $Q_t$ depends on the structure of the noise term $\mathrm{d}\omega_t$ in Eq.~\eqref{outline:sde}, see below for details.  The level 2.5 theory states that the  joint distribution of empirical measure and empirical fluxes behaves as
\beq
{\rm Prob}\left[ (\mu_t,Q_t) \approx (\mu,Q) \right] \asymp \exp[ -t I_{2.5}(\mu,Q) ]
\label{outline:L2.5}
\eeq
where $I_{2.5}$ is an explicit rate function.  The notation here a shorthand which indicates that the random variables $(\mu_t,Q_t)$ should lie inside small sets that contain the values $(\mu,Q)$, and that the equality is valid on the exponential scale as $t\to\infty$.  For a rigorous mathematical formulation of LD principles, see for example~\cite{denHollander}.

Two central results of this paper (following \cite{carollo2019}) are explicit formulae for $I_{2.5}$ for the two classes of Markovian open quantum system that were introduced in Sec.~\ref{outline:scope}. 
These results generalise existing results for large deviations at level 2.5 in classical Markov processes \cite{Maes2008,Maes2009,Barato2015,Bertini2015b,Hoppenau2016,Bertini2018}.

Large deviation principles (LDPs) at level 2.5 have several applications.  Two of the most important are: (i) they give a variational characterisation of large deviations at level-1; and (ii) they allow derivation of general bounds on fluctuations, such as thermodynamic uncertainty relations, see for example \cite{Barato2015b,Gingrich2016,PhysRevE.93.052145,Garrahan2017,Gingrich2017,Pietzonka2017,Barato_2018,PhysRevLett.125.050601,Niggemann:2020aa}.  
The connection to level 1 is discussed in detail below; the connection to thermodynamic uncertainty relations was discussed in~\cite{carollo2019}, with a brief recap in Sec.~\ref{sec:quantum-doob}, see also Appendix~\ref{app:qu-reset}.

%

\subsection{Example two-state systems}

\begin{figure}[t]
\includegraphics[width=10cm]{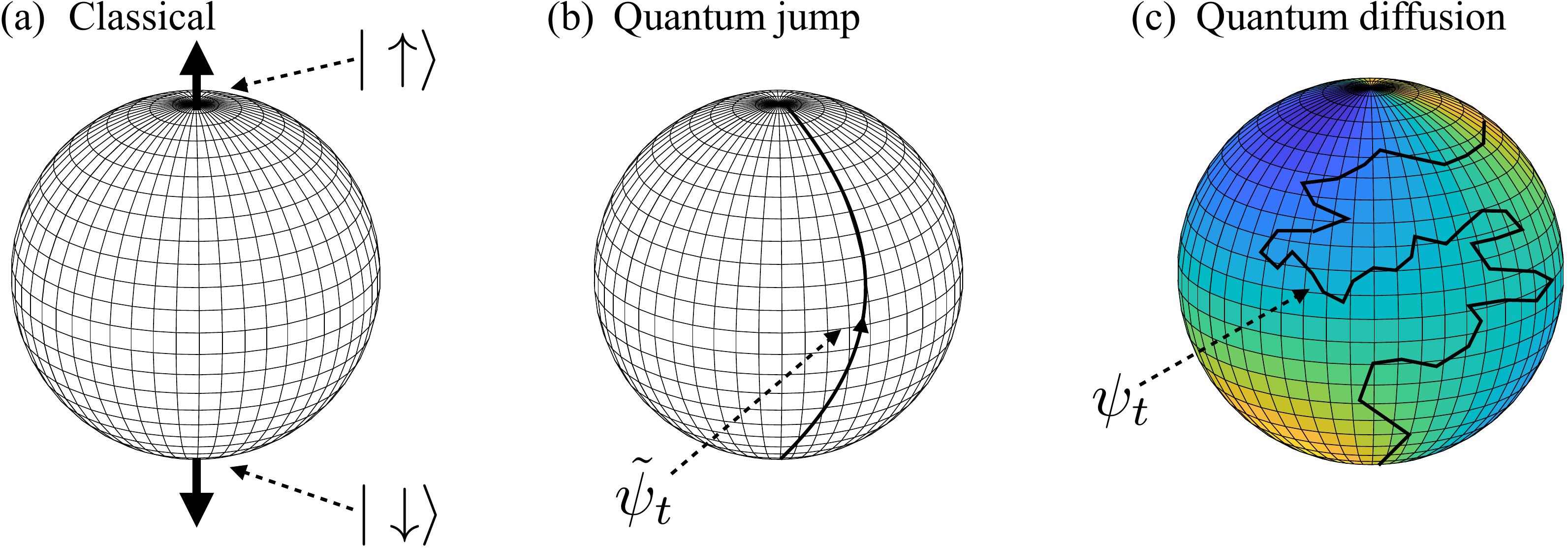}
\caption{Sketches of different empirical measures for a two level system. (a) In the classical case, where superposition is not possible, the allowed states are only the classical spin configurations with spin pointing up $|\uparrow\rangle$ (north pole) and with spin pointing down $|\downarrow\rangle$ (south pole). Therefore, the empirical measure must be given by Dirac deltas in these two points. (b) In the quantum jump example the state is reset, at every jump, to the south pole and covers a deterministic path $\tilde \psi_t$ until it jumps again. The empirical measure can thus be solely supported on such a path.  (c) In the quantum state diffusion example, stochastic trajectories are supported on the surface of the Bloch sphere and thus the empirical measure is a function defined over it. }
\label{Fig2}
\end{figure}

We illustrate the abstract arguments so far by a simple two-state quantum system, that is $n=2$.  This might represent a single spin or a single qubit.  
We emphasise that none of our results are restricted to this case, but it is useful for illustrative purposes because it allows a simple representation of the empirical objects $\mu,Q$.
In this case the most general pure-state density matrix can be represented as
\begin{equation}
\psi = \begin{pmatrix} \cos^2 \frac{\theta}{2}& \frac{1}{2}e^{-i\phi} \sin \theta \cr  \frac{1}{2}e^{i\phi} \sin \theta& \sin^2\frac{\theta}{2} \end{pmatrix}\, ,
\label{pol-coord-state}
\end{equation}
where $(\theta,\phi)$ are the spherical polar coordinates of a point on the Bloch sphere.
This means that the empirical measure $\mu$ can be interpreted as a probability distribution on this sphere.
We briefly describe three types of two-state system, in preparation for the discussion in the rest of the paper.

In a classical two-state system the only possibilities for $\psi$ correspond to the poles of the Bloch sphere, which are $\theta=0$ and $\theta=\pi$. 
Trajectories for $\psi$ are restricted to the poles [hence $b(\psi)=0$ in (\ref{outline:sde})], and they make discrete jumps from pole to pole, with randomly distributed times.  In this case, the empirical measure $\mu$ always  consists of two delta functions at the poles, with weights that indicate the time spent there, see sketch in Fig.~\ref{Fig2}(a).  The empirical flux $Q$ is a vector containing two numbers, which are the number of transitions from north to south, and the corresponding number from south to north.
The large deviations of $\mu,Q$ can be derived from the classical theory for Markov chains at level 2.5 \cite{Maes2008,Maes2009,Barato2015,Bertini2015b,Hoppenau2016,Bertini2018}.

As a second example we consider a two-state open quantum system where a light source drives transitions between the states, and there is incoherent radiative decay from state  $|\!\!\uparrow\rangle$ to state $|\!\!\downarrow\rangle$.   This corresponds to (\ref{outline:lindblad-op}) with $H=\Omega\sigma_1$, with $\sigma_1$ being the first Pauli matrix, $M=1$, and $L_1=\sigma^-\sqrt\gamma$  (with $\sigma^-=|\!\!\downarrow\rangle\langle \uparrow \!\!|$) which is also the system pictorially represented in Fig.~\ref{Fig1}.
In this case we explain below that $\mu$ is supported on a single line on the Bloch sphere, which corresponds to a deterministic evolution given by an effective (non-hermitian) Hamiltonian, starting from the south pole, as in Fig.~\ref{Fig2}(b).  The empirical flux $Q$ is a function over the path, parametrised in $t$ by $\tilde \psi_t$ indicated in Fig.~\ref{Fig2}(b), and provides the rates with which the state of the sytems at the different points in the path has jumped back to the south pole. These jumps correspond to  radiative decay events.

Finally, our third example is a two-state open quantum system coupled to a homodyne detector. We consider a fully dissipative dynamics with jump operators $L_j=i\sigma_j$, with $j=1,2,3$, proportional to Pauli matrices. The average dynamics is known as the fully dephasing channel, however we discuss how single diffusion trajectories sustain non-zero average coherences at stationarity. In this case Eq.~\eqref{outline:sde} corresponds to diffusion motion of $\psi$ on the Bloch sphere, and the empirical measure is defined over the whole sphere [see Fig.~\ref{Fig2}(c)].   In contrast to the (relatively) simple cases considered so far, the empirical flux $Q$ in this case is a more complicated object: it is related to the empirical current for the spherical diffusion. It turns out, however, that for homodyne quantum trajectories it is also necessary to introduce empirical characterizations of the noises. Details will be discussed below.

%

\subsection{Structure of the paper}
\label{subsec:structure}

Having set the scene, we outline the structure of what follows.  The statistics of quantum jumps are considered in Sec.~\ref{sec:jump}, and those of homodyne currents are discussed in Sec.~\ref{sec:diff}.  

These two main Sections are similar in structure: after introductory material in Secs.~\ref{sec:notation} and \ref{sec:unravel-jump} (respectively Secs.~\ref{sec:class-diff} and \ref{sec:diff-unravel}), the level-2.5 LD principles are presented in Secs.~\ref{sec:2.5-jump} and Sec.~\ref{sec:2.5-diff}.
Then, Sec.~\ref{sec:jump-L1-L2.5} discusses the relationships between the level-2.5 LD principle and previous results for quantum jumps at level-1, including the quantum Doob transform of~\cite{Garrahan2010,carollo2018}.  An analogous discussion is given in Sec.~\ref{sec:diff-L1-L2.5} for homodyne currents.  An example system with statistics of homodyne currents is discussed in Sec.~\ref{sec:diff-example}.

 Some of the material of Sec.~\ref{sec:jump} was presented in a shorter form in~\cite{carollo2019}, while that of Sec.~\ref{sec:diff} is original.  Compared to~\cite{carollo2019}, the discussion of Sec.~\ref{sec:jump} is more comprehensive, particularly in regard to the connections between level-2.5 and level-1, and the similarities and differences between the Doob process of the unravelled system and the quantum Doob transform~\cite{Garrahan2010,carollo2018}.  The parallel presentation of Sec.~\ref{sec:jump} and Sec.~\ref{sec:diff} emphasises the general structure of the theory.

\section{Quantum Jump Processes}
\label{sec:jump}

This section discusses the LD properties of quantum stochastic processes where the quantum state makes discontinuous random jumps. For example, such processes can describe experiments where a system emits photons that are detected by some measurement apparatus.  When photons are detected, one infers that the system has made a transition into its ground state. A shorter account of these results was presented in Ref.~\cite{carollo2019}, we also review some material from Ref.~\cite{Garrahan2010}.

\subsection{Pure-state density matrices and their calculus}
\label{sec:notation}

We introduce notation that will be important in the following.
Recall that $\psi_t$ is the pure-state density matrix of the system at time $t$.  This is a Hermitian $n\times n$ matrix.
Denote the set of all Hermitian $n\times n$ matrices by $\cal M$, and the set of pure-state density matrices by ${\cal M}_{\rm p}$ (clearly ${\cal M}_{\rm p}\subset \cal M$).  A generic member of ${\cal M}_{\rm p}$ has matrix elements 
\beq
\psi_{jk} = z^*_j z_k
\eeq
where $(z_j)_{j=1}^n$ is a (state) vector with complex elements and $\sum_{j=1}^n z_j^* z_j=1$.  (The notation $z^*$ indicates the complex conjugate of $z$.) For stochastic processes evolving mixed-state density matrices, the relevant space of states would be that of positive, unit-trace density matrices, $\mathcal{M}_{\rm m}\subset \mathcal{M}$.

The theory that we present is independent of the basis in which the pure state $\psi$ is represented.  
However, it is natural to identify a set of classical basis states $\{ |j\rangle \}_{j=1}^n$ so that $|j\rangle$ corresponds to a state vector with $z_j=1$ and  $z_k=0$ for $k\neq j$.   The corresponding matrix $\psi$ has $\psi_{jj}=1$ and all other elements are zero, it may be represented as $\psi = |j\rangle\langle j|$.

Note also that (\ref{outline:mu}) includes a (Dirac) delta function for the matrix $\psi$.  To deal with this we must define
integrals over such matrices.  It is also useful to define gradients in ${\cal M}$.  
We achieve this by treating each matrix element as a separate variable, see Appendix~\ref{app:calculus}.
For a scalar function $f=f(\psi)$ [that is, $f\colon{\cal M}\to\mathbb{R}$] 
the gradient is a matrix $\nabla f$ with elements
\beq
(\nabla f)_{jk} = \frac{\partial f}{\partial \psi_{jk}}  \; .
\label{equ:grad}
\eeq  
Also, given a matrix $X\in{\cal M}$ we define  
\beq
X\cdot \nabla f = \sum_{jk} X_{jk} (\nabla f)_{jk} \, .
\label{equ:dot}
\eeq

The theory required for integration is also outlined in  Appendix~\ref{app:calculus}.  All integrals with respect to $\psi$ (or $\psi')$ are taken over ${\cal M}$ (although in most cases the integrand is non-zero only on ${\cal M}_p$).  The key results are 
\beq
\int d\psi \, \delta(\psi-\chi) f(\psi) = f(\chi) 
\label{equ:delta}
\eeq
and an integration-by-parts formula
\beq
\int d\psi \, X(\psi) \cdot \nabla f(\psi) = - \int d\psi f(\psi) \nabla \cdot X(\psi) 
\label{equ:parts}
\eeq  
where $X=X(\psi)$ is a matrix-valued function (that is, $X\colon{\cal M}\to{\cal M}$) and its divergence is 
$\nabla \cdot X = \sum_{jk} (\partial X_{jk}/\partial \psi_{jk})$.  
Since the integrations cover the whole of ${\cal M}$, there are no boundary terms in (\ref{equ:parts}).

\subsection{Unravelled stochastic dynamics of open quantum systems, and quantum jump trajectories} 
\label{sec:unravel-jump}

We now explain how the unravelled quantum dynamics \eqref{outline:sde} operates in systems with quantum jumps.
Recalling Fig.~\ref{Fig1}, we identify 
each measurement time-record with a quantum trajectory, which specifies the state of the system at each time $t$, conditioned on the measurement outcomes obtained.  An example of a measurement time-record is the detection plot in Fig.~\ref{Fig1}(a). 
The time records -- and hence the quantum trajectories -- are generated by a stochastic process, 
described by the Belavkin equation \cite{Belavkin1990}
\begin{equation}
d{\psi}_t=\mathcal{B}[{\psi}_t]dt+\sum_i\left(\frac{\mathcal{J}_i({\psi}_t)}{\Tr[\mathcal{J}_i({\psi}_t)]}-{\psi}_t\right)dn_{it}\, ,
\label{BLS}
\end{equation}
which is the unravelled equation \eqref{outline:sde}, specialised to the jump case. 
Here, $d\psi_t$ represents the increment of the pure quantum state  $\psi_t$ in the infinitesimal time interval $[t,t+dt]$  
and 
\begin{equation}
\mathcal{B}[\psi]=-iH_{\rm eff}\psi+i\psi H^\dagger_{\rm eff}-\psi\Tr(-iH_{\rm eff}\psi+i\psi H^\dagger_{\rm eff})\, ,
\label{equ:cal-B}
\end{equation}
with 
\beq
H_{\rm eff}=H-i/2\sum_iL^\dagger_iL_i , \qquad \hbox{and} \qquad  \mathcal{J}_i[\psi]=L_i\psi L^\dagger_i \; .
\eeq
The $dn_{it}$ in \eqref{BLS} are random noise increments 
whose possible values are $0,1$; they account for the detection events, see Fig.~\ref{Fig1}(a).  Only one event can occur in any infinitesimal time period, which means that $dn_{it}dn_{jt}=\delta_{ij}dn_{it}$; the average noise increment, conditioned on the state being in $\psi_t$, is $\mathbb{E}_{\psi_t}[dn_{it}]=\Tr\left(\mathcal{J}_i[\psi_t]\right) dt$. 
We emphasise that Eq.~\eqref{BLS} describes the time-evolution of a matrix and must be interpreted as a set of equations for increments of matrix elements $(d\psi_{t})_{jk}$. 

\subsubsection{Comparison of quantum and classical processes}
\label{sec:qu-class}

Eq.~\ref{BLS} can describe both classical and quantum jump processes, on an equal footing.  
The relevant classical processes are Markov jump processes over the $n$ classical basis states.  
They are specified by transition rates $W(x,y)$  [from classical state $x$ to the classical state $y$].
Their trajectories are piecewise-constant: 
the system remains in a classical state for a (random) time interval before making a discrete jump to some other (classical) state.
Hence the allowed values of $\psi_t$ are the (discrete) classical states $|j\rangle\langle j|$ with $1\leq j \leq n$.

To describe the trajectories of these models one first sets $H=0$ in \eqref{BLS}. Then, for every non-zero rate one introduces a jump operator $
L_{xy}=\sqrt{W(x,y)}|y\rangle \langle x|$. This jump operator generates jumps of $\psi_t$ from $|x\rangle\langle x|$ to $|y\rangle\langle y|$, with rate $W(x,y)$. [The indices $i$ in \eqref{BLS} are replaced by indices $xy$, which label the types of jump.] With these conditions, by starting from a classical configuration state one has a classical state at every later time and  ${\cal B}[\psi]=0$. 

Quantum jump processes differ in several important respects from classical processes.   First, $\psi_t$ can include quantum superpositions as well as classical states: this means that $\psi_t$ can take any value from the set ${\cal M}_{\rm p}$.  Second, trajectories for $\psi_t$ are piecewise-continuous instead of piecewise constant.  In fact, the trajectories are piecewise-deterministic: $\psi_t$ evolves between jumps as $(\partial\psi_t/\partial t)=\mathcal{B}[{\psi}_t]$ which may be solved as
\begin{equation}
\psi_{t+\Delta t}=\frac{e^{-i \Delta t H_{\rm eff}}\, \psi_t \, e^{i \Delta t H_{\rm eff}^\dagger }}
  {\Tr\left[ e^{-i \Delta t H_{\rm eff}}\, \psi_t \, e^{i \Delta t H_{\rm eff}^\dagger } \right]}\, .
 \label{equ:psi-det}
\end{equation}
 The jumps are discrete, as in the classical case.  If a jump occurs at time $t$ via the $i$th jump operator then the density matrix jumps as
 \beq
\psi_t \longrightarrow \frac{\mathcal{J}_i[\psi_t]}{\Tr\left(\mathcal{J}_i[\psi_t]\right)}\, .
\label{equ:jump-dest}
\eeq
This means in particular that while classical jumps occur between discrete configurations, quantum jumps can occur between generic quantum superpositions.  Given a system in state $\psi$, the jump rate into $\psi'$ (by channel $i$) is
\begin{equation}
w_i(\psi,\psi')=\Tr\left(\mathcal{J}_i[\psi]\right)\delta\left(\psi'-\frac{\mathcal{J}_i[\psi]}{\Tr\left(\mathcal{J}_i[\psi]\right)}\right)\, .
\label{wrates}
\end{equation}
The $\delta$ function indicates that the final point of a jump is fully determined by the initial point and the channel.

The fact that the quantum state evolves continuously between jumps also has consequences for the statistics of the times at which the jumps take place. 
In particular,  the probability density function of times between jumps is exponentially distributed in classical jump processes but has a more general structure in quantum systems.

\subsubsection{Unravelled quantum master equation}

As discussed in Ref. \cite{carollo2019}, it is useful to derive a dynamical generator that describes the evolution of the quantum state given in \eqref{BLS}.  
(The relevant theory is that of piecewise-deterministic Markov processes~\cite{Breuer2002}.) 
The generator for this process is a linear functional: 
\begin{equation}
\mathcal{W}[f(\psi)]= \mathcal{B}[\psi] \cdot \nabla f(\psi)  + \sum_{i}\int d\psi'\, w_i(\psi,\psi')\left[f(\psi')-f(\psi)\right]\, .
\label{equ:Wf}
\end{equation}
(If $f$ is a matrix-valued function then ${\cal W}$ acts separately on each matrix element.)
The generator has the property  
\beq
\frac{d}{dt}\mathbb{E}[ f(\psi_t) ] = \mathbb{E}\left[ {\cal W}[f(\psi_t)] \right]
 \; .
 \label{equ:gen}
 \eeq  
We note from \eqref{equ:cal-B} that
\beq
{\cal B}[\psi] = -i[H,\psi] - \frac12 \sum_i \left( L^\dag_i L_i \psi + \psi L^\dag_i L_i \right) + \psi  \sum_i \Tr( L^\dag_i L_i \psi ) \; .
\eeq
Hence, taking $f(\psi)=\psi$ in \eqref{equ:Wf} we find
\beq
\mathcal{W}[ \psi ] = {\cal L}(\psi)
\label{equ:W-mean-L}
\eeq
where ${\cal L}$ is given by \eqref{outline:lindblad-op}.
This $\cal L$ is a linear operator.  Hence by \eqref{equ:gen}, the time evolution of $\rho(t)=\mathbb{E}[\psi_t]$ is given by (\ref{outline:lindblad-eq}).

To avoid any confusion associated with the notation in (\ref{equ:W-mean-L}), we discuss briefly the object ${\cal W}[\psi]$.
An alternative notation in (\ref{equ:Wf}) would be to write ${\cal W}f$ for the function obtained by operating with $\cal W$ on $f$, so the left hand side of (\ref{equ:Wf}) would be ${\cal W}f(\psi)$.  In this case one can define the identity function $e$ by $e(\psi)=\psi$ and the left hand side of (\ref{equ:W-mean-L}) would be ${\cal W}e(\psi)$.  Throughout this work, that object is denoted by ${\cal W}[\psi]$.

Physically, we have shown that averaging the pure state $\psi_t$ over the trajectories of the unravelled dynamics generates the (mixed) density matrix of the open quantum system of interest.  It is a non-trivial feature of these unravelled processes that the expectation value of $\psi$ obeys a closed equation of motion.  (The situation is similar to classical Ornstein-Uhlenbeck processes.)
 
The process \eqref{BLS} also has a master equation, which is an equation of motion for the probability density for $\psi_t$, which is denoted by $P_t(\psi)$.
For a generic function $f$,
\begin{equation}
\int d\psi\, f(\psi) \frac{d}{dt}P_t(\psi) = \frac{d}{dt}\mathbb{E}[f(\psi_t)] =\int d\psi\, P_t(\psi) {\cal W}[f(\psi)] \, .
\label{duality}
\end{equation}
Since this equation holds for all $f$, one obtains from (\ref{equ:Wf}) that
\begin{equation}
 \frac{d}{dt}{P}_t(\psi)=-\nabla \cdot \left[\mathcal{B}[\psi]P_t(\psi)\right]+\sum_i \int d\psi'\left[P_t(\psi')w_i(\psi',\psi)-P_t(\psi)w_i(\psi,\psi')     \right]\, ,
\label{UQME}
\end{equation}
which is the \emph{unravelled quantum master equation}~\cite{carollo2019}.  
We define an adjoint operator ${\cal W}^\dag$ via $\int d\psi\, f {\cal W}^\dag[p] = \int d\psi\, p {\cal W}[f]$, which should hold for all $p,f$. Hence from (\ref{duality}) we can also write $(d/dt)P_t(\psi) = {\cal W}^\dag[P_t(\psi)]$.  

Note that (\ref{outline:lindblad-eq}) is known as the quantum master equation (QME), but the \emph{unravelled} quantum master equation (\ref{UQME}) is a completely different object.  In particular, the unravelled QME describes the time-evolution of a probability density function, similar to standard master equations in the theory of stochastic processes.  The QME describes the time-evolution of a density matrix, and has a different structure from standard master equations.

\subsubsection{Steady state}

We assume throughout that the Hamiltonian and jump operators in \eqref{BLS} are such that the process converges for long times to a unique steady state.  
This means in particular that for any initial condition $P_0$, the solution of (\ref{UQME}) tends to a unique long-time limit which we denote by $P_\infty$ (see Ref.~\cite{Benoist:2019} for conditions on the uniqueness of this invariant measure for quantum Markov chains).
The linear operator ${\cal W}$ has eigenvalues which are non-positive, with at least one zero.  
Since the state space ${\cal M}_{\rm p}$ is compact, the uniqueness of the steady state means 
that the zero-eigenvector of ${\cal W}$ is unique and that all other eigenvalues have (strictly) negative real parts.
That is, ${\cal W}$ has a positive spectral gap.

The interpretation of $P_\infty$ is the probability density for $\psi_t$, in the steady state.
We also define the joint probability density $\Gamma$ for the initial and final points of quantum jumps, in the steady state.  This is
\beq
\Gamma_i(\psi,\psi') = P_\infty(\psi) w_i(\psi,\psi') \; .
\label{equ:Gamma}
\eeq
Also let $\Gamma$ be a vector whose elements are the $\Gamma_i$ (for $1\leq i \leq M$).

\subsection{LD principle at level 2.5}
\label{sec:2.5-jump}

We now formulate the level 2.5 LD principle for these systems, similar to (\ref{outline:L2.5}).
The empirical measure $\mu_\tau(\psi)$ was defined in (\ref{outline:mu}).   
It follows from (\ref{outline:mu},\ref{equ:delta}) that the trajectory-dependent quantity 
\beq
\int \mathrm{d}\psi f(\psi) \mu_\tau(\psi) = \frac{1}{\tau} \int_0^\tau dt f(\psi_t)
\label{equ:emp-time-ave}
\eeq 
is the empirical time-average of $f$.  
We now define the quantity that  plays the role of $Q$ in (\ref{outline:L2.5}).  
This is a vector of empirical jump rates, denoted by $k_\tau$.  
For a given trajectory, the empirical jump rate for channel $i$ depends on the initial and final points of every jump in the trajectory; it is defined by
\begin{equation}
k^i_\tau(\psi,\psi')=\frac{1}{\tau}\sum_{\hbox{\scriptsize jumps $j$ by channel $i$} }\delta(\psi_j^--\psi)\delta(\psi_j^+ -\psi') \;
\label{empiricalfluxes}
\end{equation}
where the sum is over all the quantum jumps of type (channel) $i$ that occur in the trajectory; the $j$th jump is from $\psi_j^-$ to $\psi_j^+$.  
Similarly to (\ref{equ:emp-time-ave}), integrals involving $k^i_\tau$ generate weighted sums over the jumps: for any function $g(\psi,\psi')$ then 
\beq
\int {d}\psi d\psi' g(\psi,\psi') k_\tau^i(\psi,\psi') = \frac{1}{\tau}\sum_{\hbox{\scriptsize jumps $j$ by channel $i$} } g(\psi_j^-,\psi_j^+)
\; .
\label{equ:emp-sum}
\eeq


\subsubsection{Statement of LD principle}

Since the system has a unique steady state and ${\cal W}$ has a positive spectral gap, it follows that weighted sums of the form (\ref{equ:emp-sum}) converge for large times to fixed (deterministic) values, as do time averages of the form (\ref{equ:emp-time-ave}).  This can be summarised as follows: for $\tau\to\infty$  then
\beq
(\mu_\tau,k_\tau) \to (P_\infty,\Gamma)
\label{equ:mu-k-typ}
\eeq
with probability one (see also~\cite{Barato2015}).

The LD theory describes rare events where this convergence fails. 
We state the relevant LD principle before sketching its derivation. The LD principle states that as $\tau\to\infty$ then the joint distribution of $(\mu,k)$ behaves as
\begin{equation}
{\rm Prob}[\mu,k]\asymp \exp\left( -\tau I_{2.5}^{\rm qu}[\mu,k] \right) \, .
\label{equ:Prob-mu-k-2.5}
\end{equation}
[This notation has the same meaning as (\ref{outline:L2.5}), the left hand side is to be interpreted as the probability distribution for $\mu_\tau,k_\tau$.]

From (\ref{equ:mu-k-typ}) one must have $I_{2.5}^{\rm qu}[P_\infty,\Gamma]=0$.  Fixing $(\mu_\tau,k_\tau)$ specifies the values of all
quantities of the form (\ref{equ:emp-time-ave},\ref{equ:emp-sum}).  
This means that the level 2.5 LD principle encodes the (joint) large deviation statistics of all such quantities.
The function $I_{2.5}^{\rm qu}$
is finite only if the current and flux obey a continuity condition
\begin{equation}
\nabla \cdot \left[\mathcal{B}[\psi]\mu(\psi)\right]=\sum_i\int d\psi' \left[k^i(\psi',\psi)-k^i(\psi,\psi')\right]\, .
\label{CES}
\end{equation}
Assuming that this condition holds (and that $\mu$ is a properly-normalised empirical measure) one has
\begin{equation}
I_{2.5}^{\rm qu}[\mu,k]=\sum_i \int d\psi d\psi'\, {\rm D}\Big[k^i(\psi,\psi') \Big| \mu(\psi)w_i(\psi,\psi')\Big]
\label{I25qu}
\end{equation}
where we have introduced the function 
\begin{equation}
{\rm D}[x|y]=x\log(x/y)-x+y\, .
\label{equ:Dxy}
\end{equation}
Equations (\ref{equ:Prob-mu-k-2.5}-\ref{equ:Dxy}) fully specify the level 2.5 LD principle for quantum jump trajectories.
If the continuity equation \eqref{CES} does not hold then we set formally $I_{2.5}^{\rm qu}[\mu,k]=+\infty$, this means that $(-1/\tau) \log {\rm Prob}[\mu_\tau,k_\tau]$ diverges as $\tau\to\infty$.

\subsubsection{Derivation of LD principle}
\label{sec:LD-recipe-jump}

All LD principles in this work are derived by the same general method, based on the G\"artner-Ellis theorem~\cite{denHollander,Touchette2009}.  We first define a moment-generating function (or functional) for the quantity of interest.  In this case we consider the empirical measure and flux so we define a generating functional:
\beq
G_\tau[u_1,u_2] = \mathbb{E}\left[ \exp\left( -\tau\int d\psi u_1(\psi) \mu_\tau(\psi) - \tau\sum_i \int d\psi d\psi' \, u_2^i(\psi,\psi') k^i_\tau(\psi,\psi') \right) \right]
\label{equ:Gfu}
\eeq
where $u_1\colon {\cal M}_{\rm p}\to\mathbb{R}$ is a function conjugate to $\mu$ and similarly $u_2\colon {\cal M}_{\rm p} 
\times {\cal M}_{\rm p} \to\mathbb{R}^m$ is conjugate to $k$.   The corresponding scaled cumulant generating functional (SCGF) is
\beq
\Theta[u_1,u_2] = \lim_{\tau\to\infty} \frac{1}{\tau} \log G_\tau[u_1,u_2]  \; .
\label{equ:Theta-u}
\eeq
Then by the G\"artner-Ellis theorem one has (modulo some technical assumptions that are always satisfied in the following):
\beq
I_{2.5}^{\rm qu}[\mu,k] =  \!\sup_{u_1,u_2} \!\left\{ - \Theta[u_1,u_2] -\!\int \!d\psi\, u_1(\psi) \mu(\psi) - \sum_i \!\int \!d\psi d\psi' \!u_2^i(\psi,\psi') k^i(\psi,\psi')   \right\}\, .
\label{equ:I2.5-jump-legendre}
\eeq
Moreover, we show in Appendix~\ref{app:jump-tilt} that $\Theta[u_1,u_2]$ may be characterised~\cite{carollo2019} as the largest eigenvalue of a tilted generator which is a deformed version of ${\cal W}$ in (\ref{equ:Wf}):
\begin{multline}
\mathcal{W}_u[f(\psi)]= \mathcal{B}[\psi] \cdot \nabla f(\psi)   - u_1(\psi) f(\psi) 
\\ 
+ \sum_{i}\int d\psi'\,  w_i(\psi,\psi')\left[ e^{-u_2^i(\psi,\psi')} f(\psi')-f(\psi)\right]
 \, .
 \label{equ:Wuf}
\end{multline}
For many large deviation problems, finding the largest eigenvalue of the tilted generator is prohibitively difficult.  However, a key feature of level 2.5 is that the maximisation in (\ref{equ:I2.5-jump-legendre}) can be solved in closed form, yielding (\ref{CES},\ref{I25qu}).  
This computation is described in Appendix~\ref{app:jump-ratefn}, it proceeds similarly to that of~\cite{Barato2015}.

\subsubsection{Comparison with level 2.5 for classical systems}
It is useful to compare the LD principle (\ref{equ:Prob-mu-k-2.5}) with corresponding results for classical Markov chains~ \cite{Maes2008,Barato2015}, For classical systems as described in Sec.~\ref{sec:qu-class}, the empirical jump rate (by channel $xy$) is simply
\beq
k^{xy}_\tau(\psi,\psi')= Q_\tau(x,y) \delta(\psi-|x\rangle \langle x|)\delta(\psi'-|y\rangle \langle y|) \, .
\eeq
where $Q_\tau(x,y)$ is the (classical) empirical jump rate:  the number of jumps from the classical state $x$ to the classical state $y$, normalised by $\tau$.
The corresponding jump rate (\ref{wrates}) is 
\beq
w_{xy}(\psi,\psi')=W(x,y)\delta(\psi-|x\rangle \langle x|)\delta(\psi'-|y\rangle \langle y|)\,.
\eeq
Also, the empirical measure $\mu$ is non-zero only for classical configurations: $\mu(\psi)=\sum_x\delta(\psi-|x\rangle \langle x|)\mu_{\rm cl}(x)$
where $\mu_{\rm cl}$ is the classical empirical measure, normalised as $\sum_x \mu_{\rm cl}(x)=1$.
Substituting these facts into $I_{2.5}^{\rm qu}(\mu,k)$ gives 
\beq
I_{2.5}^{\rm qu}(\mu,k)=\sum_{x\neq y} \left(Q(x,y)\log \frac{Q(x,y)}{\mu_{\rm cl}(x)W(x,y)}-Q(x,y)+\mu_{\rm cl}(x)W(x,y)\right)\, ,
\eeq
which indeed coincides with the classical level 2.5 functional \cite{Maes2008,Barato2015}.  (The sum runs over pairs of states for which $W(x,y)\neq 0$.)

To summarise: in the quantum formalism described here, classical jump processes correspond to piecewise constant trajectories for $\psi_t$, which takes values from a discrete set.  In such cases (\ref{I25qu}) becomes the classical LD principle at level 2.5.  The quantum case is more general because $\psi_t$ follows piecewise-continuous trajectories and can take any value in ${\cal M}_{\rm p}$.

\subsubsection{Auxiliary process (Doob transform, optimally-controlled process)}
\label{sec:doob-jump-unravel}

In LD theory, the rate function specifies the probability of rare events.  It is also important to characterise the \emph{mechanism} of these events -- that is,  the behaviour of trajectories with non-typical values of $(\mu_\tau,k_\tau)$.  The general LD theory explains that these (rare) trajectories can be characterised as \emph{typical} trajectories of a different system, which we call here the \emph{auxiliary process}.  This Section characterises the auxiliary process associated with the LD result (\ref{equ:Prob-mu-k-2.5}).

The derivation is related to a Doob transform and to optimal-control theory, see for example~\cite{Chetrite2015b,jack2019ergodicity}.  Note however: the auxiliary process that we describe here is associated to trajectories of the unravelled system, described by a Belavkin equation similar to~\eqref{BLS}.  This is different from the quantum Doob process discussed in~\cite{Garrahan2010,carollo2018}.  We return to this distinction in Sec.~\ref{sec:quantum-doob} below.

There is a general recipe for identifying auxiliary processes, using the tilted generator~\cite{Chetrite2015}.  For any such generator, we define the \emph{dominant} eigenfunction as the eigenfunction corresponding to the largest eigenvalue.   We focus on the tilted generator ${\cal W}_u$, and let $f_R=f_R(\psi)$ be its dominant eigenfunction.  Then the generator of the auxiliary process operates on functions $f$ as
\beq
{\cal W}^A_u[f(\psi)] = f_R(\psi)^{-1} {\cal W}_u[ f(\psi) f_R(\psi) ] - \Theta[u_1,u_2] f(\psi) \; .
\label{equ:W-Doob-u}
\eeq
For ${\cal W}^A_u$ to be a generator of a stochastic process, we require that its largest eigenvalue is zero and that the constant function $f(\psi)=1$ is the associated eigenvector: ${\cal W}^A_u[1]=0$.  This is easily verified for (\ref{equ:W-Doob-u}).
Indeed, this equation allows the auxiliary process to be constructed, dependent on $u_1,u_2$ and the associated eigenfunction $f_R$.   The generator of the auxiliary process has the same form as (\ref{equ:Wf}), but with the rates $w_i$ replaced by auxiliary rates $w^A(\psi,\psi')$.
To find the values of these rates associated to any given $(\mu,k)$ requires determination of the $u_1,u_2$ that achieve the maximum in \eqref{equ:I2.5-jump-legendre}.  
This computation can be performed, formulae for $w^A$ are given in (\ref{equ:wA-fR}) of Appendix~\ref{app:jump-ratefn}.
However, the final outcome of the computation can be obtained by direct physical reasoning, as we now explain.  

By definition of the auxiliary process, the empirical jump rates $k$ and the empirical measure $\mu$ are typical of its steady state.  
This means in particular that the mean jump rate from $\psi$ to $\psi'$ must be
\beq
w^A_i(\psi,\psi') = \frac{k^i(\psi,\psi')}{\mu(\psi)} \; .
\label{equ:wA-k-mu}
\eeq
This result fully specifies the auxiliary process for large deviations at level 2.5.  
It also gives a physical interpretation of the continuity constraint \eqref{CES}: the UQME for the auxiliary process is obtained by replacing $w$ by $w^A$ in (\ref{UQME}).  Then \eqref{CES} says that $P_t=\mu$ must be a steady state of that equation, consistent with $\mu$ being the steady state of the auxiliary process.

It is also notable that 
\beq
I_{2.5}[\mu,k] = \sum_i\int d\psi d\psi' \mu(\psi){\rm D}\Big[w_i^A(\psi,\psi')\Big|w_i(\psi,\psi')\Big] \, 
\label{equ:I2.5-wA}
\eeq
This measures the difference between the auxiliary rates and the original rates of the model.  It states that the magnitude of the rate function is determined by the amount by which the rates $w$ must be modified, in order to arrive at a model with the relevant $(\mu,k)$.

\subsection{Full counting statistics of quantum jumps (LDs at level-1)}
\label{sec:jump-L1-L2.5}

Since the level 2.5 LD principle encodes the probability for large fluctuations of \emph{all} time-averaged quantities, it can be used to recover the statistics of total quantum jump rates, which are called full counting statistics.  We show this explicitly, to indicate how the level 2.5 analysis can be applied. 
The total (empirical) jump rate for channel $i$ is obtained by integrating the empirical rate $k^i$ over all initial and final states
\begin{equation}
\bar{k}^i=\int d\psi d\psi' k^i(\psi,\psi')\, ,
\label{ph-count}
\end{equation}
This jump rate obeys a level-1 LD principle, which has been derived in previous work~\cite{Garrahan2010,carollo2018} using methods based on tilted Lindblad operators.  

This Section shows that the same result can be obtained by contraction from the level-2.5 LD principle, it also explores the relationships between the tilted Lindblad approach and the level-2.5 method described in this work.  Specifically, we review the tilted Lindblad method in Sec.~\ref{sec:jump-spectral}, after which Sec.~\ref{sec:unravel-fcs} shows that the same result can be derived from the level 2.5 LD principle.  The relationships between the methods are discussed in Sec.~\ref{sec:quantum-doob}, with a focus on the auxiliary process and the quantum Doob process.

\subsubsection{Tilted operator approach}
\label{sec:jump-spectral}
From (\ref{empiricalfluxes}), the integral (\ref{ph-count}) is the total number of jumps occurring by channel $i$ in the whole trajectory, normalised by $\tau$.
Also let $\bar k=(\bar{k}^1,\bar{k}^2,\dots,\bar{k}^M)$.
For long observation times $\tau$, the probability distribution of this observable obeys a LD principle 
$$
\mathrm{Prob}(\bar{k})\asymp \exp\left[-\tau I_1(\bar{k}) \right]\, .
$$

To show this, we follow again the general recipe of Sec.~\ref{sec:LD-recipe-jump}.  The SCGF is
\beq
\theta_k(\lambda) = \lim_{\tau\to\infty} \frac{1}{\tau} \log \mathbb{E}\left[ e^{-\tau \sum_i \lambda_i \bar{k}^i_\tau} \right]
\label{equ:theta-k}
\eeq
where ${\lambda}=(\lambda_1,\lambda_2,\dots,\lambda_M)$ is a vector of parameters conjugate to $\bar{k}$.
The SCGF may be characterised~\cite{Garrahan2010} as the largest eigenvalue of a linear operator acting on matrices $X\in {\cal M}$:
\begin{equation}
\mathcal{L}^\dagger_{{\lambda}}(X)=i[H,X]+\sum_i \left({\rm e}^{-\lambda_i} L^\dagger_i X\, L_i -\frac{1}{2}\left[ X L^\dagger_i L_i + L^\dagger_i L_i X \right]\right)\, ,
\label{TLO}
\end{equation}
For $\lambda=0$ one recovers ${\cal L}^\dag$, which is the adjoint of the operator ${\cal L}$ defined in (\ref{outline:lindblad-op}).
(This adjoint is defined by the property that $\Tr[X{\cal L}(\rho)]=\Tr[\rho{\cal L}^\dag(X)]$ for all Hermitian matrices $X,\rho$.)
Then 
$I_1(\bar{k})$ can be obtained by Legendre transform \cite{Touchette2009,Chetrite2013,Chetrite2015,Chetrite2015b}
\begin{equation}
I_1(\bar{k})=\sup_{{\lambda}}\left[-\bar{k}\cdot {\lambda}-\theta_k({\lambda})\right]\, .
\label{THEPHI}
\end{equation}
(In contrast to level 2.5, neither the SCGF $\theta_k$ nor the rate function $I_1$ can be obtained in closed form.)

\subsubsection{Level 1 full-counting statistics from the unravelled dynamics}
\label{sec:unravel-fcs}

We now give a different analysis of full-counting statistics, using the unravelled quantum dynamics~\eqref{BLS}.
The idea is to characterise the SCGF $\theta_k$ as the largest eigenvalue of a (tilted) generator for the unravelled system, similar to (\ref{equ:Wuf}).
Note that the SCGF $\theta_k$ in (\ref{equ:theta-k}) coincides with $\Theta[u_1,u_2]$ in (\ref{equ:Theta-u}) if we take $u_1=0$ and $u_2^i(\psi,\psi')=\lambda_i$.
Using (\ref{equ:Wuf}), it follows that $\theta_k$ can be characterised as the largest eigenvalue of the tilted generator
\beq
\mathcal{W}_\lambda[f(\psi)]= \mathcal{B}[\psi] \cdot \nabla f(\psi)    
+ \sum_{i}\int d\psi'\,  w_i(\psi,\psi')\left[ e^{-\lambda_i} f(\psi')-f(\psi)\right]
 \, .
 \label{equ:W-lambda}
\eeq

We now show explicitly that solving this eigenproblem for $\theta_k$ is equivalent to finding the largest eigenvalue of (\ref{TLO}).  
To this end,  
we first show that if $\theta$ is (any) eigenvalue of ${\cal L}^\dag_\lambda$ then it is also an eigenvalue of $\mathcal{W}_\lambda$.  In this case we have ${\cal L}^\dag_\lambda(\ell) = \theta\ell$, where $\ell$ is the relevant eigenmatrix.  Now define $f_\ell(\psi) = \Tr(\ell\psi)$. 
Since ${\cal W}_\lambda$ is a linear operator we have ${\cal W}_\lambda[ f_\ell(\psi) ] = \Tr(\ell {\cal W}_\lambda[\psi] )$.  Also, it is easily shown [by analogy with (\ref{equ:W-mean-L})] that
\beq
{\cal W}_\lambda[\psi] = {\cal L}_\lambda(\psi) \; .
\label{equ:W-lambda-L}
\eeq
so that 
\beq
{\cal W}_\lambda[ f_\ell(\psi) ] =
\Tr\left[ \psi\left( {\cal L}_\lambda^\dag(\ell) \right) \right] = \theta f_\ell(\psi)
\label{equ:Tr-L-theta-modified}
\eeq
where the second equality uses that $\ell$ is an eigenmatrix of ${\cal L}^\dag_\lambda$, and the definition of $f_\ell$.  Hence this $f_\ell$ is an eigenfunction for ${\cal W}_\lambda$ with eigenvalue $\theta$.  However the converse does not hold: there may be eigenvalues of ${\cal W}_\lambda$ that are not eigenvalues of ${\cal L}^\dag_\lambda$.  

It therefore remains to show that the largest eigenvalue of ${\cal W}_\lambda$ coincides with the largest eigenvalue of ${\cal L}_\lambda^\dag$.  For a general linear operator, we refer to the eigenfunction corresponding to the largest eigenvalue as the \emph{dominant eigenfunction}.  From our assumption that \eqref{BLS} has a unique steady state, it follows that the dominant eigenfunction of ${\cal W}_\lambda$ is always positive, $f(\psi)>0$, and that this property is unique to the dominant eigenfunction.   Moreover, the theory of Lindblad operators~\cite{carollo2018} shows that the dominant eigenmatrix $\ell$ of ${\cal L}_\lambda$ has positive eigenvalues.   Since $\psi$ is a pure state ($\psi=|z\rangle\langle z|$) then this implies $f_\ell(\psi)=\Tr(\ell\psi)=\langle z|\ell|z\rangle>0$.  So $f_\ell(\psi)$ is an eigenfunction of ${\cal W}_\lambda$ that is always positive -- it must be the dominant eigenfunction.
Hence the largest eigenvalues of ${\cal L}_\lambda$ and ${\cal W}_\lambda$ are both equal to $\theta_k$.  The level-1 rate function can then be obtained from \eqref{THEPHI}.

Finally, we observe one more way of characterising $I_1$.
By the contraction principle for LDs~\cite{denHollander,Touchette2009},
one has
\begin{equation}
I_1(\bar{k})=\inf_{\mu,k | \bar{k} } I_{2.5}^{\rm qu}[\mu,k]\, ,
\label{equ:I1-contract}
\end{equation}
where the infimum is taken over $(\mu,k)$, subject to (\ref{ph-count}).
Admissible choices for $\mu,k$ in (\ref{equ:I1-contract}) also require that $\mu$ is normalised and that the continuity condition \eqref{CES} holds.
This minimisation was performed in~\cite{carollo2019}, which verified that it is equivalent to \eqref{THEPHI}.  However, the approach here based on the tilted generator ${\cal W}_\lambda$ is a more direct route to the same answer.

\subsubsection{Auxiliary process and quantum Doob process}
\label{sec:quantum-doob}

\begin{figure}
\includegraphics[width=7cm]{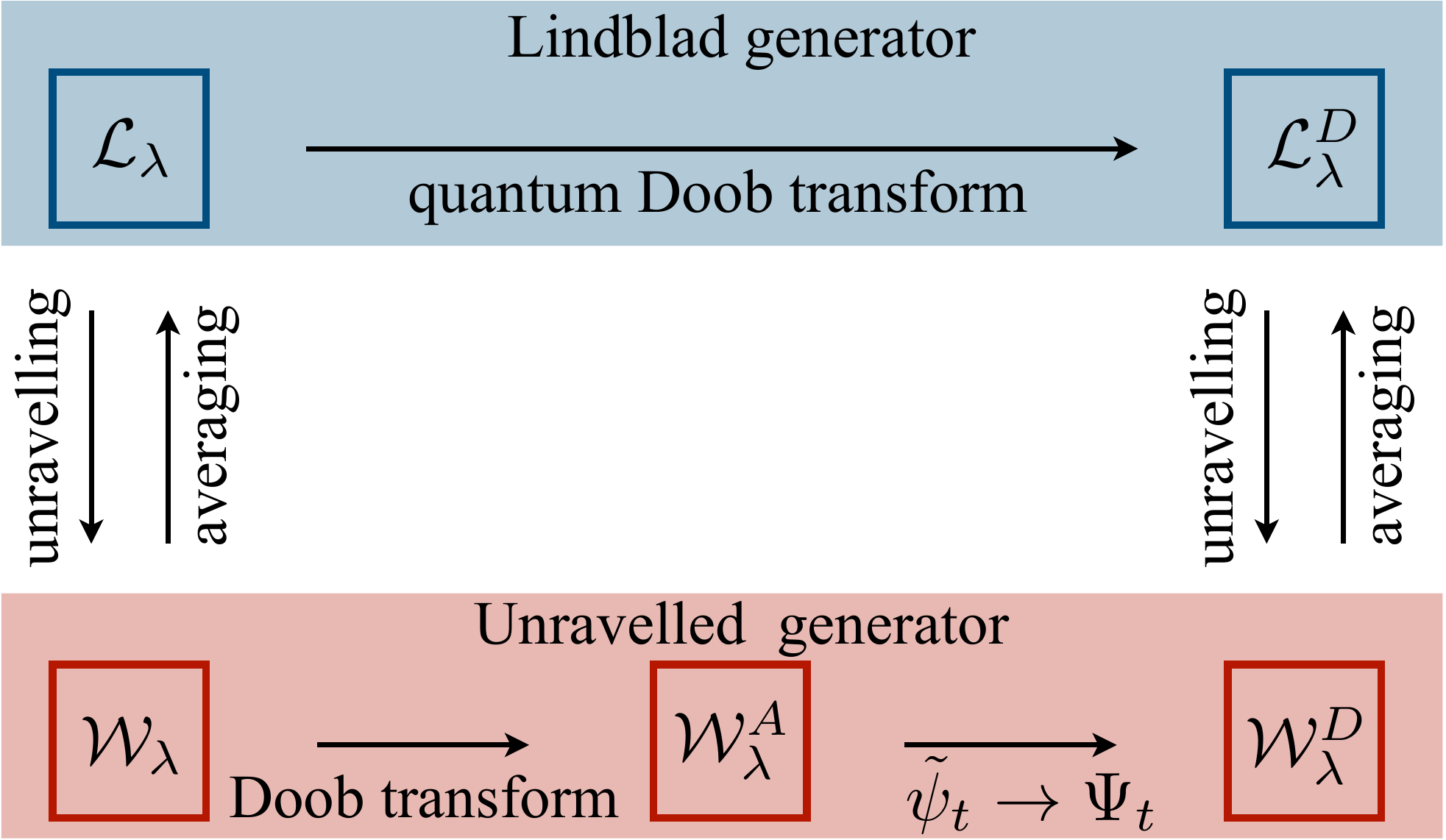}
\caption{ Illustration of the relationships between the tilted generators ${\cal L}_\lambda$ and ${\cal W}_\lambda$; the quantum Doob process (described by Lindblad generator ${\cal L}^{D}_\lambda$); and the unravelled auxiliary processes $\tilde\psi_t$ and $\Psi_t$ (described by classical generators ${\cal W}_\lambda^{A}$ and ${\cal W}_\lambda^{D}$).  It is notable that averaging the auxiliary process $\tilde\psi_t$ does not yield a valid Lindblad evolution for $\mathbb{E}[\tilde\psi_t]$.  However, the transformation (\ref{equ:Psi}) yields an unravelled process $\Psi_t$ that is related to the quantum Doob by $\mathbb{E}[\Psi_t]=\rho^{D}$, see Appendix~\ref{app:doob-mean}.}
\label{fig:doob}
\end{figure}

We now turn to the auxiliary process for full-counting statistics, which illustrates the physical connection of the unravelled dynamics to the quantum Doob process of~\cite{Garrahan2010,carollo2018}, and hence to the tilted Lindblad operator.  (The connections are summarized in Fig.~\ref{fig:doob}, below.)

In contrast to the level 2.5 LD principle where explicit results were available, LD results at level-1 rely on the solution to the eigenproblems discussed above.
However, the auxiliary rates $w^A$ are available from (\ref{equ:wA-fR}) [in Appendix~\ref{app:jump-ratefn}], in  terms of the dominant eigenfunction of ${\cal W}_\lambda$: they are
\beq
w^{\rm A}_i(\psi,\psi')=w_i(\psi,\psi'){\rm e}^{-\lambda_i}\frac{\Tr(\ell \, \psi')}{\Tr(\ell \, \psi)} \;.
\label{equ:wA-ell}
\eeq
The auxiliary process with these rates reproduces the rare (large deviation) trajectories of the unravelled process, as in Sec.~\ref{sec:doob-jump-unravel}.  Similar to \eqref{equ:W-Doob-u}, the generator of this auxiliary process is
\beq
{\cal W}^A_\lambda[f(\psi)] = \Tr(\ell\psi)^{-1} {\cal W}_\lambda[ f(\psi) \Tr(\ell\psi) ] - \theta_k(\lambda) f(\psi) \; .
\label{equ:W-Doob-lambda}
\eeq

In~\cite{Garrahan2010,carollo2018}, a different kind of auxiliary process was identified, which we call here the quantum Doob process.  It corresponds to a Lindblad equation of the form (\ref{outline:lindblad-eq}), where the Hamiltonian and the jump operators are both modified from the original model of interest.
Specifically, the Lindblad generator of this model is given by~\cite{Garrahan2010,carollo2018}
\begin{equation}
\mathcal{L}_\lambda^D[\rho]=\ell^{1/2} \mathcal{L}_\lambda[\ell^{-1/2} \rho \ell^{-1/2}]\ell^{1/2}-\theta_k(\lambda)\rho \, .
\label{Av-Doob}
\end{equation}
Using this $\mathcal{L}_\lambda^D$ in the Lindblad evolution (\ref{outline:lindblad-eq}) defines an open quantum system 
in which the Hamiltonian $\tilde{H}$ and the jump operators $\tilde L$ depend on $\lambda$ as~\cite{Garrahan2010,carollo2018}
\begin{align}
\tilde{H} & =  \frac12 \ell^{1/2} \left( H - \frac{i}{2} \sum_i L^\dag_i L_i \right) \ell^{-1/2} + \hbox{h.c.}
\nonumber\\
\tilde{L}_i & = {\rm e}^{-\lambda_i/2} \ell^{1/2} L_i \ell^{-1/2}
\label{equ:doob-HJ}
\end{align}
where h.c. denotes the Hermitian conjugate.  (We recall that $\ell$ depends on $\lambda$.)
This new system is the quantum Doob process. 
It is significant because \emph{typical} time-records of quantum jumps in the quantum Doob process match exactly the \emph{rare} time-records that appear as large deviations in the original system.   In this sense, the quantum Doob process plays the same role as the auxiliary process for the unravelled dynamics.

The unravelled dynamics for the quantum Doob process may also be constructed.  Let the pure-state density matrix of the auxiliary process of (\ref{equ:W-Doob-lambda}) be $\tilde\psi_t$ and define a new pure-state density matrix:
\beq
\Psi_t=\frac{\ell^{1/2} \tilde\psi_t\ell^{1/2}}{\Tr\left(\ell \tilde\psi_t\right)}\, .
\label{equ:Psi}
\eeq
The trajectories of this $\Psi_t$ define an unravelled jump process which was shown in~\cite{carollo2019} to coincide with the unravelled dynamics of the quantum Doob process.
This is verified in Appendix~\ref{app:doob-mean}.
In other words, the unravelled dynamics of the quantum Doob system can be obtained by deforming the auxiliary process derived here, according to \eqref{equ:Psi}.  The generator for this unravelled dynamics is denoted by ${\cal W}^D_\lambda$, it can be constructed by analogy with (\ref{equ:Wf}), with the transformed Hamiltonian and jump operators from (\ref{equ:doob-HJ}) used in place of the original $H,L$.

The relationships between the quantum Doob process and the various unravelled process are illustrated in Fig.~\ref{fig:doob}.  It is notable that the auxiliary process described by (\ref{equ:wA-ell},\ref{equ:W-Doob-lambda}) cannot generically be interpreted as the unravelled dynamics of a system obeying Lindblad dynamics (\ref{outline:lindblad-eq},\ref{outline:lindblad-op}).  (The Lindblad form places constraints on the unravelled dynamics which are not satisfied by generic auxiliary processes.)  The transformation (\ref{equ:Psi}) is essential for relating the unravelled auxiliary processes to the quantum Doob transform.

An application of the level-2.5 LD principle for quantum jumps was considered in~\cite{carollo2019}, which derived a thermodynamic uncertainty relation for photon counts, in the restricted setting of quantum reset processes.  An expanded version of that derivation is given in Appendix~\ref{app:qu-reset}.

\subsubsection{Other LDs at level-1}

So far we have considered level-1 LDs of $\bar k$, which are full-counting statistics.  
These can be investigated either using the unravelled dynamics (via ${\cal W}_\lambda$) or by a tilted Lindblad operator ${\cal L}_\lambda$.
However, working with the unravelled dynamics allows other LD principles at level-1, which cannot be obtained by tilted Lindblad methods.

To see this, consider the fluctuations of a function $\mathcal{O}=\mathcal{O}(\psi_t)$. Its time-average is 
\beq
o_\tau= \frac{1}{\tau} \int_0^\tau dt\,  \mathcal{O}(\psi_t)\, ,
\eeq
In typical cases of interest, the function $\mathcal{O}(\psi)$ might be the quantum expectation of an operator $X$, e.g. $\mathcal{O}(\psi)=\Tr\left(X\psi\right)$, which is a linear function of $\psi$.  Non-linear functions can also be considered: for example, large deviations of the entanglement entropy of a bipartite quantum system were considered in~\cite{carollo2019}.

The probability density for $o_\tau$ obeys a LD principle with 
\beq
{\rm Prob}(o_\tau)\asymp e^{-\tau \phi(o_\tau)}\, ,
\eeq
where $\phi(o)$ is the LD rate function.
Similar to Sec.~\ref{sec:unravel-fcs}, this function may be obtained by contraction from level 2.5.  Alternatively the SCGF for $o_\tau$ is $\Theta[u_1,u_2]$ from (\ref{equ:Theta-u}) with $u_1(\psi)=\lambda{\cal O}(\psi)$ and $u_2=0$.  Hence this SCGF can be obtained as the largest eigenvalue of the appropriate operator ${\cal W}_u$ and the rate function can be obtained by Legendre transform (with $\lambda$ as the conjugate field). Furthermore, it is also possible to estimate this SCGF by using population dynamics methods \cite{Giardina2006,Giardina2011} applied to the unravelled master equation \eqref{UQME} \cite{PhysRevE.102.030104}. We are not aware of any general characterisation of such SCGFs in terms of tilted Lindblad operators.

\section{Quantum Diffusion Processes}
\label{sec:diff}

 Many stochastic processes in classical physics are described by differential equations involving Wiener noises (or Langevin equations, or Brownian motions).  In the large deviation context, these processes also obey LD principles at level 2.5. The ideas are similar to jump processes, but the technical details are different.  In particular the empirical current plays the role of the flux $Q$ in jump processes.

In the quantum context, homodyne measurements on open quantum systems result in random output signals that are related to Brownian motions, recall Fig.~\ref{Fig1}.  (This is in contrast the photon-detection experiments which are related to jump processes.)

We emphasize that the presentation of this Section is analogous to Sec.~\ref{sec:jump}, with the addition of an example system that is analysed in Sec.~\ref{sec:diff-example}.  The LD principle at level-2.5 is presented in Sec.~\ref{sec:2.5-diff} and the connection to level 1 is discussed in Sec.~\ref{sec:diff-L1-L2.5}, including the relation between quantum Doob process and unravelled auxiliary process.  To set up those results,
we briefly review level 2.5 functionals for classical diffusion processes \cite{Barato2015,Chetrite2015b}, and we explain how these are generalised to the quantum case of homodyne detection experiments \cite{Gardiner2004}.

\subsection{Summary of LDs at level 2.5 for classical diffusion processes}
\label{sec:class-diff}

As a generic classical diffusion process we take $x\in\mathbb{R}^d$ evolving by a stochastic differential equation with $n_\alpha$ independent noises:
\begin{equation}
dx_t=A(x_t)dt +\sum_{\alpha=1}^{n_\alpha} B_\alpha dW_t^\alpha\, ,
\label{Class-Diff}
\end{equation}
where $A\colon \mathbb{R}^d \to \mathbb{R}^d$ is a drift term, and $B_\alpha \in \mathbb{R}^d$ is a vector indicating the strength and direction of noise $\alpha$ (assumed independent of $x_t$). The $W^\alpha$ are independent Wiener processes with unit variance.
We define a diffusion matrix with elements
\beq
D^{ij}_{\rm cl}=\frac12 \sum_\alpha 
B_\alpha^i B_\alpha^j \, ,
\eeq
for $1\leq i,j\leq d$.
Here and elsewhere, sums over $\alpha$ are assumed to run from $1$ to $n_\alpha$.
The matrix $D_{\rm cl}$ is assumed to be invertible \cite{Barato2015} which requires (as a necessary condition) that $n_\alpha\geq d$.
We assume that this model has a unique steady state.

In this section, the natural geometry is that of Euclidean space $\mathbb{R}^d$: gradients such as $\nabla f$ and dot products such as $A\cdot\nabla f$ are taken in this space.  [In later sections we revert to gradients in $\cal M$, as defined in (\ref{equ:grad}).]

The generator for (\ref{Class-Diff}) is ${\cal W}_{\rm diff}$ which acts on functions $f\colon \mathbb{R}^d\to\mathbb{R}$ as
\begin{equation}
{\cal W}_{\rm diff}[f]
=A\cdot \nabla f+\sum_{ij}D_{\rm cl}^{ij}\frac{\partial^2 f}{\partial x_i\partial x_j}\, .
\label{Diff-Markov}
\end{equation}
Sums over $i,j$ are taken over $1,2,\dots,d$.
Similar to (\ref{duality}), the generator can be used to derive the Fokker-Planck equation for the time-evolution of the probability density for $x_t$:
\beq
\dot{P}_t=-\nabla \cdot J_{\rm cl}(P_t)
\label{equ:classical-FP}
\eeq
where $J_{\rm cl}$ is the probability current which depends linearly on $P_t$. Its elements are
\begin{equation}
[J_{\rm cl}(P_t)]_i=A_i P_t-\sum_{j}D_{\rm cl}^{ij}\frac{\partial P_t}{\partial x_j}\, .
\label{stat-current}
\end{equation}
In the steady state one has $P_t=P_\infty$ and the associated probability current $J_{{\rm cl},\infty}=J_{\rm cl}(P_\infty)$ is divergence-free: $\nabla \cdot J_{{\rm cl},\infty}=0$.

\subsubsection{Large deviations}

To analyse large deviations at level 2.5, we define the empirical measure using (\ref{outline:mu}), as above.  We also define an empirical current as
\beq
{J}^{e}_\tau(x)=\frac{1}{\tau}\int_0^\tau \delta (x_t-x)\circ dx_t \, ,
\label{equ:Je-classical}
\eeq
where the $\circ$ symbol indicates that the integral uses the Stratonovich convention.  
For a given trajectory, $J^e_\tau(x)$ measures the displacement of the system at point $x$, summed over its visits to that point, and divided by the total time.

For large times we have a result analogous to (\ref{equ:mu-k-typ}), which is
\beq
(\mu_\tau,J^e_\tau) \to (P_\infty,J_{{\rm cl},\infty}) 
\label{equ:mu-Je-typ}
\eeq
with probability one.
The corresponding LD principle is
$$
{\rm Prob} [(\mu_\tau,J^e_\tau)\approx(\mu,J)]\asymp \exp\left[-\tau I_{2.5}^{\rm cl}(\mu,J) \right] \, , 
$$
which describes the joint statistics of empirical measure and empirical current.  The associated rate function is finite only if $\nabla \cdot J=0$, in which case it takes the value
\begin{equation}
I_{2.5}^{\rm cl}(\mu,J)=\frac{1}{4}\int dx \left[ J-J_{\rm cl}(\mu) \right] \cdot (\mu D_{\rm cl})^{-1} \left[ J-J_{\rm cl}(\mu) \right] \,  ,
\label{class-diff-func}
\end{equation}
as discussed in~\cite{Barato2015,Chetrite2015b}.

\subsubsection{Empirical noise}
Note the presence of the inverse of $D_{\rm cl}$ in (\ref{class-diff-func}), so this matrix should be invertible to apply the theory as presented here.  
For quantum diffusions, the analogue of this matrix may not be invertible.  This motivates us to modify the standard theory at level 2.5, as follows.  We define an \emph{empirical noise}
\beq
j^\alpha_\tau(x) =\frac{1}{\tau} \int_0^\tau \delta(x_t-x) dW_t^\alpha
\eeq
which is the average noise increment for particles at $x$.  
(Note, this integral is taken in the Ito sense, so the empirical noise has mean zero.)
Writing $j$ for the vector of empirical noises, it can be shown that the distribution of $(\mu_t,j_t)$ obeys an LDP [similar to (\ref{equ:Prob-mu-k-2.5})]
\beq
{\rm Prob} (\mu,j)\asymp \exp\left[-\tau I_{\rm noise}^{\rm cl}(\mu,j) \right] \, . 
\eeq 
(We only state this result here, we give the corresponding derivation for the quantum case below.
That derivation is easily adapted to this case.)
The rate function $I_{\rm noise}^{\rm cl}$ is finite only if a suitable continuity condition $\nabla \cdot J=0$ holds; this can also be written as 
\beq
\nabla \cdot \left[ A \mu + \sum_\alpha B_\alpha j^\alpha - D_{\rm cl} \nabla\mu  \right] = 0 \; .
\label{equ:CES-classical}
\eeq
The object inside square brackets is the empirical current $J$ for a system with empirical measure $\mu$ and noise $j$. It is the sum of $J_{\rm cl}(\mu)$ and a term coming from the empirical noise.
In cases where the continuity condition (\ref{equ:CES-classical}) holds, the rate function is simply
\beq
I_{\rm noise}^{\rm cl}(\mu,j)  = \frac12  \int dx \sum_\alpha \frac{ j^\alpha(x)^2 }{ \mu(x) }
\label{equ:I-noise}
\eeq
The level 2.5 rate function (\ref{class-diff-func}) can be obtained from this LD principle by contraction: one minimises $I^{\rm cl}_{\rm noise}$ over all empirical noises $j^\alpha$ that are consistent with a given empirical current.  This minimisation yields the inverse of $D_{\rm cl}$ in cases where it exists.  
(We note in passing that these discussions are not mathematically rigorous, in particular we have not stated precise technical conditions required on (\ref{Class-Diff}) in order to obtain this LD principle, although we do insist that the system should have a unique steady state \cite{Benoist:2019}.  See also the discussion of the quantum case, below.)

Physically, the meaning of this contraction is that large deviations occur via the \emph{least unlikely} noise realisations, and $j_\tau$ characterises these noises.   In cases where $D_{\rm cl}$ does not have an inverse, there are some empirical currents that cannot be realised by \emph{any} realisation of the noise.  In this case the quantity  $J - J_{\rm cl}(\mu)$ in (\ref{class-diff-func}) is outside the image of $D_{\rm cl}$ and the rate function $I_{2.5}^{\rm cl}$ is formally infinite.

Note finally, there is a thermodynamic uncertainty principle for currents in these systems~\cite{Gingrich2017-jpa,nardini2018}, it is straightforwardly derived by setting $\mu=P_\infty$ and $J=\lambda J_{\rm cl}(P_\infty)$ in \eqref{class-diff-func}, which manifestly solves \eqref{equ:CES-classical}.  This construction is valid only if $D_{\rm cl}$ has the property that $J_{\rm cl} = D_{\rm cl} F_{\rm cl}$ may be solved for $F_{\rm cl}$ (for all $x$ where $P_\infty>0$).  The simplest case is when $D_{\rm cl}^{-1}$ exists but it is sufficient in general that the steady-state current $J_{\rm cl}(P_\infty)$ can be represented as $\sum_\alpha B_\alpha j^\alpha $, so that there exist realisations of the empirical noise that generate a uniform acceleration of the steady-state current.

\subsection{Homodyne detection experiments: unravelled dynamics}
\label{sec:diff-unravel}

In the case of homodyne detection experiments, quantum systems are monitored by a continuous observation of the quadratures of the bath quantum operators. This type of measurement process allows a  detailed characterization of the emissions of the system into the environment \cite{Hickey2012}. Indeed, since quadrature operators are proportional to the intensity of the light field, the measured homodyne current can provide information not only about the overall number of emitted photons, but also about the nature of the light, i.e. whether this is in a thermal, coherent or more complex state. In ideal conditions, the outcome of a homodyne experiment consists of a record of the time-integrated value of the measured current as a function of time, as shown in Fig.~\ref{Fig1}(b). These time-records are stochastic and depend both on the quantum state and on the specific realization of the noisy interaction between system and environment. In what follows we present a comprehensive discussion of the large deviations in these systems starting from the level 1 statistics for homodyne currents and then deriving the very general level 2.5 functional encoding the statistics of generic observable of the process.

Throughout this section, we consider a system described by the Lindblad evolution (\ref{outline:lindblad-eq},\ref{outline:lindblad-op}),  as for jump processes.  However, we slightly change our notation in that jump operators are labelled by $m$ (with $1\leq m \leq M$) instead of by $i$.

\subsubsection{Stochastic Schr\"odinger equation}

To describe homodyne trajectories, we consider a stochastic Schr\"odinger equation as in \eqref{outline:sde}. This takes the form of an (Ito) stochastic differential equation similar to (\ref{Class-Diff}):
\begin{equation}
d \psi_t= \mathcal{L}(\psi_t)dt +\sum_{m=1}^M \mathcal{K}^m(\psi_t) dW_m\, ,
\label{QSSE}
\end{equation}
where ${\cal L}$ is the Lindblad operator from (\ref{outline:lindblad-op}) and 
\begin{align}
\mathcal{K}^m(\psi)&=\kappa_m(\psi)-\psi\Tr\left[\kappa_m(\psi)\right]\, ,
\nonumber \\
\kappa_m(\psi)&=e^{i\alpha_m}L_m\psi+\psi e^{-i\alpha_m}L_m^\dagger\, ,
\label{unrav-diff-kappa}
\end{align}
where $\alpha_m$ is a phase factor (see below) and the $L_m$ are the jump operators appearing in (\ref{outline:lindblad-op}).
In contrast to (\ref{Class-Diff}), the noise strengths ${\cal K}^m$ depend on the state $\psi$.  This means that we must take care to use Ito's formula when evaluating increments of $\psi$-dependent functions.  In the literature on quantum diffusions~\cite{Gardiner2004}, this is implemented by Ito rules
\begin{equation}
\mathbb{E}[dW_m]=0, \qquad \mathbb{E}[dW_m dW_n]=\delta_{mn}dt\, . 
\label{Wiener}
\end{equation}
The phases $\alpha_m$ in (\ref{unrav-diff-kappa}) specify the particular quadrature operator of the environment modes that the experiment is monitoring \cite{Hickey2012}, for each homodyne current.  The Lindblad evolution (\ref{outline:lindblad-eq}) is independent of these phases but the unravelled trajectories can depend qualitatively on the $\alpha_m$.

Equation \eqref{QSSE} is a stochastic differential equation which describes every possible time-record of a homodyne experiment in which the state is being continuously monitored. In particular, a typical outcome consists of the values of the time-integrated homodyne currents $Q^m$.  These are random (trajectory-dependent) quantities, given by  
\begin{equation}
Q^m_\tau=\int^\tau_0 d Q^m_t,\quad  \mbox{with}\quad d Q^m_t=\Tr\left[\kappa_m(\psi_t)\right]dt +dW_m\, .
\label{HomoCurr}
\end{equation}
Let $Q_\tau$ be a vector whose elements are the $Q^m_\tau$.  

Comparing  \eqref{QSSE} with (\ref{Class-Diff}) one sees that $\cal L$ describes the drift of the diffusion process while ${\cal K}$ describes the noises.  From the Ito rules (\ref{Wiener}) one sees immediately that $\mathbb{E}(d\psi_t) = \mathbb{E}({\cal L}(\psi_t)) dt$; using that ${\cal L}$ is a linear operator yields $\mathbb{E}(d\psi_t) = {\cal L}(\mathbb{E}(\psi_t)) dt$.   Recalling that $\mathbb{E}(\psi_t)=\rho_t$ is the density matrix, one recovers (\ref{outline:lindblad-eq}).  That is, the fact that the drift term is linear in the Ito equation (\ref{QSSE}) means the expectation value of $\psi$ obeys a closed (linear) equation. (The same is true for Ornstein-Uhlenbeck equations in the classical setting.)

Unless otherwise stated, we assume in the following that the unravelled process has a unique steady state in which the probability density for $\psi_t$ is $P_\infty(\psi)$, as in the case of quantum jump processes.

\subsubsection{Unravelled quantum master equation}

The next step is to identify the generator for the stochastic process (\ref{QSSE}).  
We compute this at the level of the quantum state $\psi$. 
Consider a function $f=f(\psi$): its increment $df$ in the short time interval $[t,t+dt]$ is obtained by Taylor-expanding to second order:
\beq
d f = \sum_{ij}\frac{\partial f}{\partial \psi_{ij}}  (d \psi_t)_{ij}+\frac{1}{2}\sum_{ij,hk}\frac{\partial^2 f}{\partial \psi_{ij} \partial \psi_{hk}}(d \psi_t)_{ij}(d \psi_t)_{hk}
 \; 
\label{Stoc-Fun}
\eeq
(It is implicit throughout this section that sums run over all allowed values of the relevant index.)
Taking the expectation and using (\ref{Wiener}) yields
\beq
\mathbb{E}[d f ]=  \mathbb{E}\left[ \sum_{ij}\frac{\partial f}{\partial \psi_{ij}} (\mathcal{L}[\psi])_{ij}
+\frac{1}{2}\sum_{ij,hk}\frac{\partial^2 f}{\partial \psi_{ij}\psi_{hk}}D_{ij,hk}(\psi) \right] dt
\label{Aveg-Fun}
\eeq
where 
\begin{equation}
D_{ij,hk}(\psi)=\sum_m\left(\mathcal{K}^m[\psi]\right)_{ij}\left(\mathcal{K}^m[\psi]\right)_{hk}\, 
\label{Diff-Equ}
\end{equation}
is the analogue of the classical diffusion matrix $D_{\rm cl}$ in this setting (up to a factor of $2$).
Following that analogy, one sees that that if the number of terms in the sum ($M$) is not large enough, the matrix $D$ will be degenerate, and the inverse $D^{-1}$ will not exist.  Indeed, this situation is likely to be common for systems under homodyne measurement. 

Using
(\ref{Aveg-Fun},\ref{equ:gen}) and recalling  (\ref{equ:dot}) we identify the generator for functions of $\psi$ as
\begin{equation}
\mathcal{W}[f(\psi)]= \mathcal{L}[\psi] \cdot \nabla f(\psi)
+\frac{1}{2}\sum_{ij,hk}D_{ij,hk}(\psi) \frac{\partial^2 f}{\partial \psi_{ij}\partial \psi_{hk}}\, ,
\label{Mark-gen}
\end{equation}
which is analogous to the classical result (\ref{Diff-Markov}).
Taking $f(\psi)=\psi$ recovers again that $\rho=\mathbb{E}(\psi)$ evolves as in (\ref{outline:lindblad-eq}).

The analogue of (\ref{equ:classical-FP}) is the \emph{unravelled quantum master equation for diffusion processes}:
\begin{equation}
\dot{P}_t=-\nabla \cdot {J}(P_t)\, , 
\label{FP}
\end{equation}
where $P_t$ is the probability density for $\psi$. The corresponding probability current is a matrix-valued function of $P$, its elements are
\begin{equation}
\left[{J}(P)\right]_{ij}=\mathcal{L}_{ij}P-\frac{1}{2}\sum_{hk}\frac{\partial}{\partial \psi_{hk}}\left(D_{ij,hk} P\right) \; .
\label{current}
\end{equation}

\subsection{LD at level 2.5 for quantum diffusions}
\label{sec:2.5-diff}

We now derive a LD principle at level 2.5, following a similar method to Sec.~\ref{sec:2.5-jump}.  
The empirical measure is given as usual by (\ref{outline:mu}).
Analogous to the classical case from Sec.~\ref{sec:class-diff}, we define the empirical noises
\beq
j^m_\tau(\psi)=\frac{1}{\tau}\int_0^\tau \delta (\psi_t-\psi)  \, dW_m\, .
\eeq
Note that
\beq
Q^m_\tau = \tau \int d\psi \left\{ \Tr[\kappa^m(\psi)] \mu(\psi) + j^m(\psi) \right\} \; .
\label{equ:Q-noise}
\eeq
The empirical current is
\beq
{J}^{\rm e}_\tau(\psi)=\frac{1}{\tau}\int_0^\tau\delta(\psi_t-\psi)\circ d\psi_t\, .
\label{equ:Je-psi}
\eeq 
Note that (\ref{equ:Je-psi}) includes a Stratonovich product, in contrast to the Ito products used elsewhere.  Taking care with this fact we show in
Appendix~\ref{app:Je-strato} that the empirical current is fully determined by the empirical measure and empirical noise, as
\beq 
[J^e_\tau(\psi)]_{ij} = \mu_\tau(\psi)[{\cal L}(\psi)]_{ij}  + \sum_m  j_\tau^m(\psi) [{\cal K}^m(\psi)]_{ij} - \frac12 \sum_{hk}  \frac{ \partial}{\partial \psi_{hk}} (\mu_\tau(\psi) D_{ij,hk}(\psi)) \; .
\label{equ:Je-j}
\eeq

\subsubsection{Large deviation principle and auxiliary dynamics}

We derive a LD principle for $(\mu_\tau,j_\tau)$ noting that large deviations of $(\mu_\tau,Q_\tau,J^e_\tau)$ can then be obtained by contraction.  To achieve this,
we follow the same steps as Secs.~\ref{sec:LD-recipe-jump} and \ref{sec:doob-jump-unravel}.  We give a short presentation of the computation, referring to those earlier sections for context and discussion.

Define a moment generating functional for $(\mu,j)$ that takes as arguments $a_1\colon {\cal M}_p\to \mathbb{R}$ and 
$a_2\colon {\cal M}_p\to \mathbb{R}^M$:
\beq
G_\tau[a_1,a_2] = \mathbb{E}\left[ \exp\left( \tau \int d\psi a_1(\psi) \mu_\tau(\psi) + \tau\sum_m \int d\psi a_2^m(\psi) j^m_\tau(\psi) \right) \right]\, .
\label{equ:scgf-qudiff}
\eeq
The corresponding SCGF is
\beq
\Theta[a_1,a_2] = \lim_{\tau\to\infty} \frac{1}{\tau} \log G_\tau[a_1,a_2]  \; .
\label{equ:Theta-a}
\eeq
The resulting LD principle is
\beq
\label{equ:diff-L2.5}
{\rm Prob} (\mu,j)\asymp \exp\left[-\tau I^{\rm qu}_{2.5}(\mu,j) \right] \, , 
\eeq
with
\beq
I_{2.5}^{\rm qu}(\mu,j) = \sup_{a_1,a_2} \left\{ \int d\psi a_1(\psi) \mu(\psi) + \sum_m \int d\psi a_2^m(\psi) j^m(\psi)  - \Theta[a_1,a_2] \right\} \; .
\label{equ:diff-2.5-sup}
\eeq
Recall, this last formula should be obtained by applying the G\"artner-Ellis theorem to \eqref{equ:Theta-a}.  
For a rigorous treatment, this would require technical conditions on $\Theta$, which we do not explore here.  From a physical perspective, we expect $\Theta$ to be well-behaved as long as the unravelled system explores its (unique) steady state within some finite mixing time \cite{Benoist:2019}.  We assume that this is the case and the G\"artner-Ellis theorem can be applied -- such a requirement is not trivial in systems where $D$ is non-invertible, but we do not expect this to be too restrictive a condition in practice.

This supremum can be computed exactly: we state the (simple) result before outlining the derivation. The rate function is finite only if a continuity equation  holds: 
the empirical current $J^e$ during large deviation events must converge, for large times, to a current $J$ which must be divergence-free, $\nabla\cdot {J} = 0$, because the relevant trajectories are stationary.
From \eqref{equ:Je-j}, this requires
\beq
\sum_{ij} \frac{\partial}{\partial \psi_{ij}} \left[ {\cal L}_{ij}\mu  + \sum_m {\cal K}^m_{ij} j^m - \frac12 \sum_{hk}  \frac{ \partial}{\partial \psi_{hk}} (\mu D_{ij,hk})   \right] = 0 \; .
\label{equ:CES-diff}
\eeq
(For compactness of notation, we omit functional dependence on $\psi$ where this leaves no ambiguity.)
In cases where \eqref{equ:CES-diff} holds then
\begin{equation}
{I}^{\rm qu}_{2.5}(\mu,j)=\frac{1}{2}\int d\psi \, \sum_{m=1}^M \frac{ j_m^2(\psi) }{ \mu(\psi) } \, .
\label{Q-funct}
\end{equation}

Just as in the classical case (Section~\ref{sec:class-diff}), a thermodynamic uncertainty relation can be derived in this system, if there exist choices of empirical noise such that $ \sum_m {\cal K}^m j^m $ in (\ref{equ:CES-diff}) is proportional to the steady state current $J(P_\infty)$.  One simply substitutes these noises in \eqref{Q-funct} with $\mu=P_\infty$ so (\ref{equ:CES-diff}) is easily satisfied.  In cases where this construction is not possible, we are not aware of any thermodynamic uncertainty relation.

To derive (\ref{equ:CES-diff},\ref{Q-funct}), we show in Appendix~\ref{app:diff-tilt} that $\Theta[a_1,a_2]$ from (\ref{equ:Theta-a}) is  the largest eigenvalue of the tilted operator
\beq
\mathcal{W}_{a}[f]=   \left[ \mathcal{L} +\sum_{m}a_2^m \mathcal{K}^m \right] \cdot \nabla f 
+\frac{1}{2}\sum_{ij,hk}D_{ij,hk}\frac{\partial^2 f}{\partial \psi_{ij} \partial \psi_{hk}}
+ a_1 f +\frac{1}{2}\sum_m (a_2^m)^2 f 
\, .
\label{Tilted-A}
\eeq
The derivation of (\ref{Q-funct}) from this operator is given in Appendix~\ref{app:diff-ratefn}, it
is similar to that of Appendix~\ref{app:jump-ratefn} for the jump case.

Similar to Sec.~\ref{sec:doob-jump-unravel}, the auxiliary dynamics is explicit for level 2.5.  
It may be derived by identifying its generator as
\beq
{\cal W}^A_a[f(\psi)] = f_R(\psi)^{-1} {\cal W}_a[ f(\psi) f_R(\psi) ] - \Theta[a_1,a_2] f(\psi) \; ,
\eeq
where $f_R$ is the dominant eigenvector of ${\cal W}_a$.  This is similar to (\ref{equ:W-Doob-u}).
Physically, the meaning of the auxiliary process is that
the noise $dW^m$ develops a ($\psi$-dependent) mean value equal to $(j^m/\mu)$.   Hence \eqref{QSSE} is modified in the auxiliary dynamics as
\beq
d \psi_t= \left[ \mathcal{L}(\psi_t) + \sum_{m} \frac{ \mathcal{K}^m(\psi_t) j^m(\psi_t) }{ \mu(\psi_t) } \right] dt +\sum_{m} \mathcal{K}^m(\psi_t) dW_m\, .
\label{equ:dpsi-cont-2.5}
\eeq
Similarly, from \eqref{HomoCurr} one sees that the homodyne current in this auxiliary model evolves as
\beq
d Q^m_t= \left[ \Tr\left[\kappa_m(\psi_t)\right] + \frac{j_m(\psi_t)}{\mu(\psi_t)} \right] dt + dW_m
\eeq
We recognise \eqref{equ:CES-diff} as the condition that this auxiliary dynamics has steady-state distribution $\mu$, similar to the discussion of Sec.~\ref{sec:doob-jump-unravel} for jump processes.

\subsection{LDs of homodyne currents (level 1)}
\label{sec:diff-L1-L2.5}

Building on this analysis of level 2.5 LDs, we now consider LDs of homodyne currents.  As in the case of jump processes, this will allow us to recover (and extend) earlier work that was based on tilted Lindblad operators~\cite{Hickey2012}.  

We define the time-averaged homodyne current as
\beq
q^m_\tau = \frac{1}{\tau} Q^m_\tau \; .
\eeq
 Similar to Sec.~\ref{sec:jump-L1-L2.5}, the level-1 LD principle for this quantity can be computed from the unravelled LD principle at level-2.5, or by using tilted Lindblad operators.  The relation between the corresponding auxiliary processes and quantum Doob processes are also analogous to Sec.~\ref{sec:jump-L1-L2.5}.  Given that the physical picture is the same as that Section, the presentation here is brief.

\subsubsection{Tilted operators}

We analyse large deviations of $q_\tau$.  Its SCGF is
\beq
\theta_q(s) = \lim_{\tau\to\infty} \frac{1}{\tau} \log \mathbb{E}\left[ \exp\left( - \tau \sum_m s_m q_\tau^m \right) \right]
\eeq
where $s=(s_1,s_2,\dots,s_M)$ is the field conjugate to $q$.
By (\ref{equ:Q-noise}), $\theta_q$ coincides with $\Theta[a_1,a_2]$ from \eqref{equ:Theta-a} with $a_1=-\sum_m s_m \Tr[\kappa^m(\psi)]$ and $a_2^m=-s_m$.
Hence it suffices to consider the largest eigenvalue of the operator
\begin{multline}
\mathcal{W}_{s}[f]=   \left[ \mathcal{L} - \sum_{m} s_m \mathcal{K}^m \right] \cdot \nabla f 
+\frac{1}{2}\sum_{ij,hk}D_{ij,hk}\frac{\partial^2 f}{\partial \psi_{ij} \partial \psi_{hk}}
\\ 
-  \sum_m s_m  \Tr[\kappa^m(\psi)] f +\frac{1}{2}\sum_m (s^m)^2 f 
\, ,
\label{Tilted-Q}
\end{multline}
As in Sec.~\ref{sec:unravel-fcs}, the dominant eigenfunction $f$ turns out to be linear in $\psi$.  Similar to (\ref{equ:W-mean-L}) we have
\beq
{\cal W}_s[\psi] = {\cal L}_s(\psi)
\label{equ:W-mean-L-s}
\eeq
with a tilted Lindblad generator
\begin{equation}
\mathcal{L}_s(\rho)=\mathcal{L}(\rho)+\sum_{m} \left[ \frac12 (s_m)^2  \rho - s_m \kappa_m(\rho) \right] \; .
\label{tilt-Lind-s}
\end{equation}
Repeating the argument of Sec.~\ref{sec:unravel-fcs}, if $\ell$ is an eigenmatrix of $ {\cal L}_s^\dag$ with eigenvalue $\ell$, then $f(\psi) = \Tr(\ell\psi)$ is an eigenfunction of ${\cal W}_s$, with the same eigenvalue.

The operator ${\cal L}_s$ is a multivariate generalisation of the tilted operator derived in \cite{Hickey2012}.  
(The definition of $s$ used here differs from theirs by a factor of 2.)
Analysis of this operator shows that the dominant eigenmatrix $\ell$ has positive eigenvalues, which means that the largest eigenvalue of ${\cal W}_s$ is
also the largest eigenvalue of ${\cal L}_s$, and therefore coincides with $\theta_q(s)$.
The operator ${\cal L}_s$ was used in \cite{Hickey2012} to analyse large deviations of homodyne currents at level-1.
Here we have shown how these large deviations can be analysed directly from the unravelled trajectories.

\subsubsection{Auxiliary process and quantum Doob process for homodyne detection}

Similar to the discussion of full-counting statistics in Sec.~\ref{sec:quantum-doob}, 
one may construct an auxiliary model that reproduces the quantum trajectories associated with large deviation events.
One may also construct a quantum Doob-transformed process similar to those in~\cite{Garrahan2010,carollo2018}.  

For the auxiliary process in the unravelled representation, one has from \eqref{equ:jm-am} (in Appendix~\ref{app:diff-ratefn}) that the empirical noise associated to the rare event is
\beq
j^m(\psi) = \mu(\psi) \left( \frac{  \Tr[\ell{\cal K}^m(\psi)]  }{ \Tr[\ell\psi] } - s_m\right)
\eeq
(We used that $a_2^m=-s_m$ and $f_R=\Tr[\ell\psi]$, note that if $s=0$ then $\ell$ is the identity and $j^m=0$, as required.)  
From \eqref{equ:dpsi-cont-2.5} the auxiliary process for $\psi$ is then
\beq
d \psi_t= \left[ \mathcal{L}(\psi_t) + \sum_{m}  \mathcal{K}^m(\psi_t) \left( \frac{  \Tr[\ell{\cal K}^m(\psi_t)]   }{ \Tr[\ell\psi_t] } - s_m\right)  \right] dt 
  +\sum_{m} \mathcal{K}^m(\psi_t) dW_m\, .
\label{equ:doob-homodyne}
\eeq

For the quantum Doob process one follows instead the procedure given in~\cite{Garrahan2010,carollo2018}.  The resulting Lindblad operator is
\begin{equation}
\mathcal{L}_s^D[X]=\ell^{1/2} \mathcal{L}_s[\ell^{-1/2} X\ell^{-1/2}]\ell^{1/2}-\theta(s)X\, .
\label{Av-Doob-homo}
\end{equation}
This corresponds to a physical model whose typical trajectories allow reconstruction of the homodyne measurement records associated with large deviations event of the original model, analogous to the case of full-counting statistics.

Following again the argument of Appendix~\ref{app:doob-mean} [using (\ref{equ:W-mean-L-s})] shows that the unravelled dynamics of the quantum Doob process can be related to that of the auxiliary process (\ref{equ:doob-homodyne}) by (\ref{equ:Psi}), just as in the case of full counting statistics.  (Recall also Fig.~\ref{fig:doob}.)

\subsection{Example}
\label{sec:diff-example}

We illustrate the application of the level 2.5 formalism with an example from a two-level quantum system, 
corresponding to a quantum spin-$1/2$ particle. The space of states in this case admits a pictorial representation in terms of the Bloch sphere. In particular, pure states are parametrised by the spherical polar coordinates $(\theta,\phi)$ as in~\eqref{pol-coord-state}.  
We show that the unravelled system corresponds to diffusion on this sphere, and we analyse large deviations in this case.   
The large deviations of the homodyne currents are quite trivial in this case, so we discuss instead large deviations of the (time-integrated) coherence of the unravelled quantum state.

We consider the dissipative Lindblad dynamics 
\begin{equation}
\dot{\rho}_t=\sum_{m=1}^3 \left(\sigma_m  \rho_t  \sigma_m- \rho_t\right)\, , 
\label{Lind-diff-example}
\end{equation}
where $\sigma_m$ is the $m$-th Pauli matrix. 
The unravelled trajectories are generated by~\eqref{QSSE}, with $\alpha_m=\pi/2$ in Eq.~\eqref{unrav-diff-kappa}. Hence
\begin{equation}
d\psi_t=\sum_{m=1}^3 \left(\sigma_m  \psi_t \sigma_m- \psi_t\right) dt +i\sum_{m=1}^3 [\sigma_m,\psi_t ]  dW_m  \, .
\label{example-stoc-diff}
\end{equation}
Note that the stochastic term resembles unitary evolution with a random time-dependent Hamiltonian given (formally) by $\sum_m \sigma_m (dW_m/dt)$.

\subsubsection{Diffusion on the Bloch sphere}

For two-state systems, it is natural to write the density matrix in spherical co-ordinates, as in~\eqref{pol-coord-state}.
The equation of motion can then be represented as
\begin{equation}
d\begin{pmatrix}
\theta_t\cr\phi_t\end{pmatrix}=2\begin{pmatrix}
\cot\theta_t\cr0
\end{pmatrix}dt+\mathcal{K}
\begin{pmatrix}
dW_1\cr dW_2\cr dW_3 \end{pmatrix} 
\label{example-stoc-diff-angles}
\end{equation}
with
\beq
\mathcal{K}=2\begin{pmatrix}
\sin\phi_t&-\cos \phi_t&0\cr
\cot \theta_t\cos\phi_t&\cot \theta_t\sin\phi_t&-1 \end{pmatrix}\, .
\eeq
This representation emphasises that the set of matrices ${\cal M}_{\rm p}$ can be parameterised by two real angles.  Hence,
instead of considering a probability density $P(\psi)$ as in previous analysis, one may consider a simple probability density for $\theta,\phi$.
This obeys a Fokker-Planck equation
\begin{equation}
\frac{\partial P_t}{\partial t}=-2\frac{\partial}{\partial \theta}\left(\cot\theta P_t\right) +2\frac{\partial^2P_t}{\partial \theta^2}+2\csc^2\theta \frac{\partial^2P_t}{\partial \phi^2}\, ;
\label{example-FP-noncov}
\end{equation}
alternatively, noting that the uniform distribution on the sphere corresponds to $P(\theta,\phi) = \sin\theta/(4\pi)$ one may define a covariant probability density $p(\theta,\phi)=P(\theta,\phi)/\sin\theta$ which evolves as
\beq
\frac{dp_t}{dt}=2\left[\frac{1}{\sin \theta}\frac{\partial }{\partial \theta}\left(\sin \theta \frac{\partial p_t}{\partial \theta}\right)+\frac{1}{\sin^2\theta }\frac{\partial^2 p_t}{\partial \phi^2}\right]\, .
\eeq
We recognise the right hand side as the Laplacian in spherical co-ordinates.  The meaning of this equation is that $\psi$ undergoes isotropic diffusion on the Bloch sphere, and the steady-state distribution is uniform on the sphere.  It follows that the steady-state of the system is time-reversal symmetric, and the probability current in this state is zero.

We work primarily with $P(\theta,\phi)$, the probability density for $(\theta,\phi)$.  In this representation, the probability current $J$ is a tangent vector to the Bloch sphere, it has components along the azimuthal ($\phi)$ and polar ($\theta$) directions.  That is
\beq
J(P) = ( J_\theta(P), J_\phi(P) )
\label{equ:J-angles}
\eeq
and from \eqref{example-FP-noncov} one has  $J_\theta = 2[\cot\theta P + (\partial P/\partial \theta)]$ and 
$J_\phi = -2 \csc^2 \theta (\partial P/\partial\phi)$.
From this point, the classical large deviation theory of Sec.~\ref{sec:class-diff} can be applied, as we see below.

We note however that this geometrical representation based on the Bloch sphere is limited to two-state quantum systems ($n=2$).  
The general theory that we have presented is based on probability densities for matrix elements of $\psi$, which we have denoted by $P(\psi)$.  
In this case the probability current is matrix-valued quantity, as in (\ref{current}).  Since such currents may not be intuitive, we give a brief physical discussion of how they appear in this case.

\subsubsection{Currents in ${\cal M}$}

The key point from \eqref{example-stoc-diff-angles} is that while $\psi$ has 4 elements (two real and two complex), a general increment of $\psi$ can be described by a two-component vector $(d\theta,d\phi)$, because $\psi$ always remains in ${\cal M}_{\rm p}$.   The current is similarly described by the two components $(J_\theta,J_\phi)$.

There is a useful geometrical structure here: to illustrate it in a simple way, we consider a smooth curve on the Bloch sphere that is described by $\psi(u)$ with $0\leq u \leq 1$.  The restriction to smooth paths (and not Brownian motions) avoids complications from Ito's formula.  The curve can be specified in terms of two functions $(\theta(u),\phi(u))$.  Then
\beq
\psi'(u) = E_\theta(\psi) \theta'(u) + E_\phi(\psi) \phi'(u) 
\label{equ:vel-EE}
\eeq
where primes indicate derivatives, and 
\beq
E_\theta(\psi) = \begin{pmatrix} -\sin\theta & \frac{1}{2}e^{-i\phi} \cos \theta \\
\frac{1}{2}e^{i\phi} \cos \theta & \sin\theta \end{pmatrix}
, \qquad
E_\phi(\psi) =  \begin{pmatrix} 0 & - \frac{i}{2}e^{-i\phi} \sin \theta \\
\frac{i}{2}e^{i\phi} \sin \theta & 0 \end{pmatrix}
\eeq 
are $\psi$-dependent matrices which form a basis for the tangent space to the Bloch sphere.  
 It follows that the probability current at the point $\psi$ is
\beq
J(\psi) = E_\theta(\psi) J_\theta(\psi) + E_\phi(\psi) J_\phi(\psi)
\eeq
where $E_\theta,E_\phi$ are the matrices from (\ref{equ:vel-EE}) and $J_\theta,J_\phi$ are the scalar fields defined on the sphere as in (\ref{equ:J-angles}).
The empirical current also has a similar form.

This general picture still holds true for systems with $n>2$ states: the probability current is a vector field that is everywhere tangent to ${\cal M}_p$
and can be written as $J = \sum_\alpha E_\alpha(\psi) J_\alpha(\psi)$ where the $E_\alpha$ are matrices that form a basis for the tangent space and the $J_\alpha$ are real-valued fields.  Explicit construction of the basis matrices $E$ and the currents $J$ is not simple for large $n$ (the tangent basis has $2(n-1)$ independent components).  This motivates our general formulation in terms of matrix elements of $\psi$, which is always applicable. 

\begin{figure}[t]
\includegraphics[width=6cm]{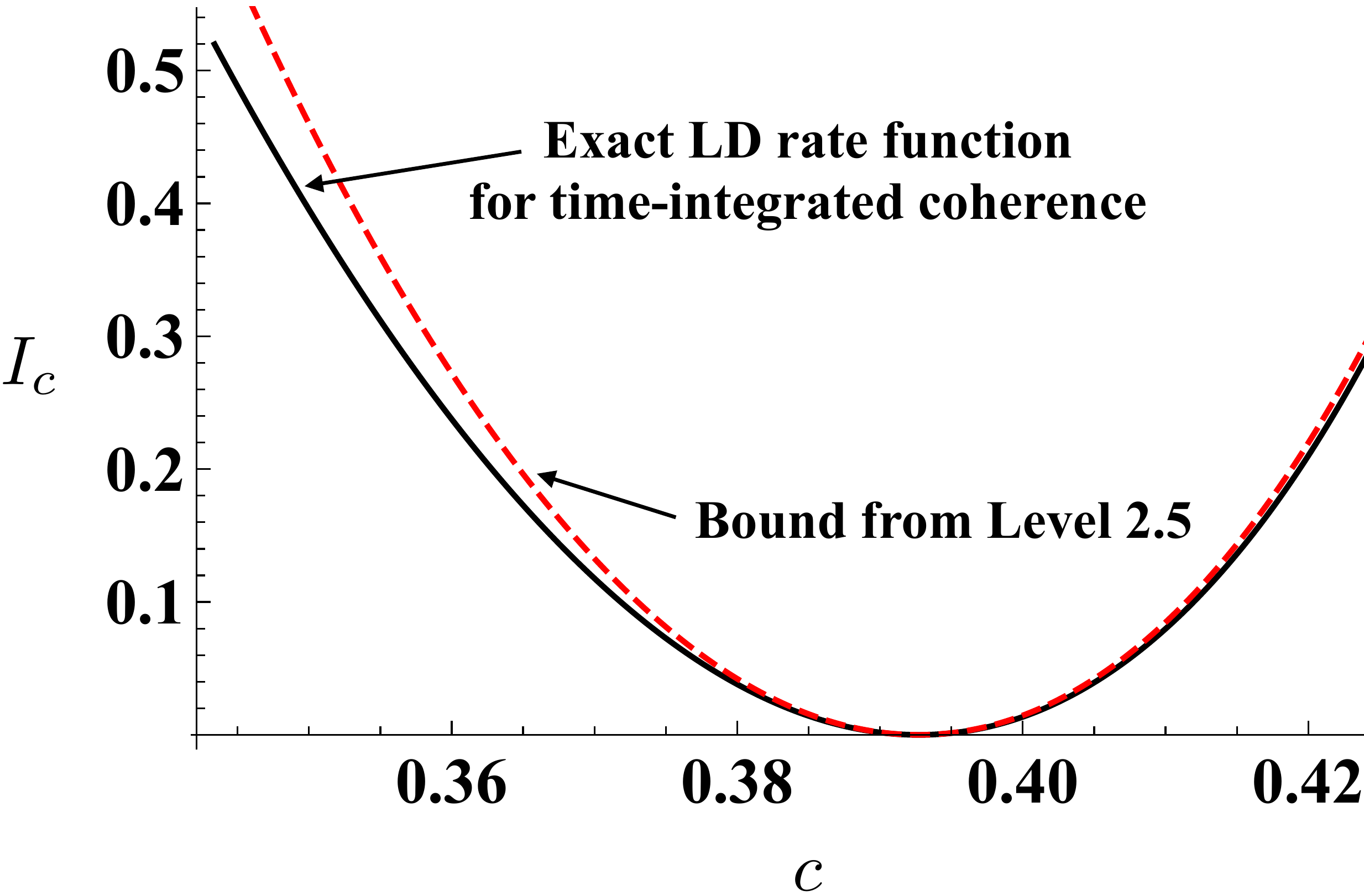}
\caption{Large deviation rate function for time-integrated coherence in the diffusion example model~\eqref{Lind-diff-example}. The black solid line is the exact LD rate function which has been obtained by Legendre transform of the scaled cumulant generating function, which is the largest real eigenvalue of the twisted Fokker-Planck operator in Eq.~\eqref{example-twisted-FP}. The dashed red line is the bound for this rate function which was been found by evalulating the level 2.5 rate function with (\ref{equ:mu-lambda-diff},\ref{equ:j-lambda-diff}). }
\label{FigBoundDiff}
\end{figure}

\subsubsection{Large deviation analysis}

Returning to the example (\ref{example-stoc-diff}), it is easily verified from (\ref{HomoCurr}) that $dQ^m = dW^m$, that is, the three homodyne currents are simple random walks.  Hence their rate functions are simple quadratic functions.  This is a general feature of systems where the operators $e^{i\alpha_m}L_m$ are anti-Hermitian.  
We note that steady state of the Lindblad evolution in our example is $\rho\propto{\bf 1}$, the identity matrix, so there are no coherences in the steady state, at this level.

However, while the homodyne currents have Gaussian fluctuations, large deviations of the empirical measure $\mu$ can have more complex behaviour, and so can coherences within the unravelled system.  We consider the coherence of the unravelled density matrix
\beq
{\cal C}(\psi) = |\psi_{12}| = (\sin\theta)/2
\eeq
This object does have non-trivial fluctuations in the unravelled dynamics.  We define its time average
\begin{equation}
{\bar c}_\tau=\frac{1}{\tau} \int_0^\tau dt \, \mathcal{C}(\psi_t)\, . 
\label{example-diff-coher}
\end{equation}
For large observation times $\tau$, one has that 
\beq
\lim_{\tau\to\infty} \mathbb{E}[{\bar c}_\tau]=\int_0^\pi d\theta \int_0^{2\pi} d\phi\, \mathcal{C}(\psi) P_\infty(\theta,\phi)=\frac{\pi}{8}\, .
\eeq
The coherence also obeys an LD principle ${\rm Prob}(\bar c_\tau) \asymp e^{-\tau I_c(\bar c_\tau)}$ with (by a contraction argument)
\beq
I_c(\bar c) = \inf_{\mu,j} I_{2.5}(\mu,j)
\label{equ:inf-Ic}
\eeq
where the infimum is subject to $\int d\psi \mu(\psi) {\cal C}(\psi) = \bar{c}$.   

The rate function $I_c$ can also be obtained as $\sup_s [ -s\bar c - \Theta_c(s) ]$ where $\Theta_c$ is the largest eigenvalue  that solves 
\begin{equation}
-2\frac{\partial}{\partial \theta}(P \cot\theta)+2\frac{\partial^2P}{\partial \theta^2}+2\csc^2\theta\frac{\partial^2P}{\partial \phi^2}+ \frac{s}{2} P \sin \theta  = \Theta_c(s) P
\label{example-twisted-FP}
\end{equation}
This problem can be solved numerically (see Fig.~\ref{FigBoundDiff}).  We now obtain a simple bound on this rate function via level 2.5.  This requires that we construct a suitable $\mu,j$ for use in (\ref{equ:inf-Ic}).

We use \eqref{equ:Je-j} to express the empirical current of our example system in terms of its empirical noises, as
\begin{align}
{J}^e_\theta&=2\mu \cot\theta -2\frac{\partial\mu}{\partial \theta} +2j_1\sin\phi -2j_2\cos\phi \, ,
\nonumber \\
{J}^e_\phi&=-2\csc^2\theta \, \frac{\partial\mu}{\partial \phi }+2j_1\cot\theta\cos\phi+2j_2\cot\theta\sin\phi-j_3\, .
\end{align}
We require $(\partial {J}^e_\theta/\partial\theta) + (\partial {J}^e_\phi/\partial\phi) = 0$ in order solve the continuity constraint \eqref{equ:CES-diff}.
In fact, the time-reversal symmetry of the original problem and the fact that the coherence is time-reversal symmetric 
means that the relevant large deviations are realised by trajectories with ${J}^e=0$. 

We express $\mu$ as a probability density for $(\theta,\phi)$, just like $P$.
We consider a one-parameter family of $(\mu,j)$, in order to generate a range of values for $\bar c$.  Specifically,
\beq
\mu_\lambda(\theta,\phi)=\frac{1}{2\pi}\frac{e^{-\lambda \sin \theta}\sin \theta}{\int_0^\pi d\theta'\, e^{-\lambda \sin \theta'}\sin \theta'}\, .
\label{equ:mu-lambda-diff}
\eeq
This is independent of $\phi$, just like $P_\infty$ (the coherence does not depend on $\phi$ so there is no reason why the auxiliary process should perturb its distribution away from that of the steady state).  For $\lambda>0$ the distribution $\mu_\lambda$ is biased towards the poles of the Bloch sphere (less coherence) and for $\lambda<0$ it biases towards the equator (more coherence).  Then suitable empirical noises that achieve $J^e=0$ are 
\begin{equation}
j_1= f(\theta) \sin \phi\, ,\qquad j_2=-f(\theta)\cos\phi\, ,\qquad j_3=0\, ,
\label{equ:j-lambda-diff}
\end{equation}
with $f(\theta) = \frac{\partial\mu}{\partial \theta}  -\mu\cot\theta$.  Physically, this empirical noise counteracts the tendency of the system to relax towards the steady state $P_\infty$, leading to trajectories with non-typical values of the coherence.  Using these $(\mu_\lambda,j)$, the value of $I_{2.5}$ can then be computed.  To provide a bound on $I_c$ one must also compute $\int d\psi \mu(\psi) {\cal C}(\psi)$.  Performing the integrals numerically, the resulting bound is shown in Fig.~\ref{FigBoundDiff}, together with the (numerically) exact result obtained via \eqref{example-twisted-FP}.

The bound reproduces the general behaviour of $I_c$ but is not exact.  The reason is that the ansatz \eqref{equ:mu-lambda-diff} is not sufficient to fully capture the empirical measure of the rare trajectories that realise the rare event.  However, it does have the right general form, especially for small $\lambda$.  The fact that the rare trajectories have $J^e=0$ is also an accurate reflection of the rare event -- one way to see this is that writing \eqref{example-twisted-FP} as an equation for the covariant density $p$ transforms the eigenvalue problem, into the (time-independent) Schrodinger equation for energy levels of a quantum particle diffusing on a sphere, with potential $(s\sin\theta)/2$.   This is a Hermitian eigenvalue problem: these correspond generically to large deviation problems with time-reversal symmetry~\cite{Garrahan2007}.

\section{Outlook}
\label{sec:outlook}

We have analysed unravelled stochastic processes that describe the dynamical evolution of open quantum systems.  In particular, we have derived LD principles for these systems at level 2.5, which provide an explicit and general characterisation of the joint fluctuations of the system and environment.  
We have explained how these LD principles are related to LDs at level 1, which can be analysed by tilted operator methods~\cite{Garrahan2010,Hickey2012,cilluffo2020microscopic}.
We have also discussed the implications of these LD principles for thermodynamic uncertainty relations.

This work opens up a framework for analysing fluctuating behaviour in open quantum systems. It provides a thermodynamic formalism where new interesting nonequilibrium phenomena, such as entanglement phase transitions \cite{Nahum2017,Nahum2018,Keyserlingk2018,PhysRevX.9.031009,PhysRevB.101.104301,ippoliti2020entanglement,alberton2020trajectory} or scrambling of information in stochastic processes \cite{PhysRevB.98.195125,PhysRevB.98.184416}, can be investigated by means of concepts and tools that proved very powerful in equilibrium statistical mechanics. This makes it possible to look not only at average behaviour but also at the behaviour of higher-order time-correlation functions in quantum stochastic processes. 

Since quantum trajectories have a direct relation with measurement outcomes in actual experimental settings involving continuously monitored systems, as for example the photon-counting or homodyne-detection experiments discussed here, this formalism also brings the theoretical investigation of nonequilibrium quantum systems closer to real observations. Given the general connection between large deviation principles and gradient-flow dynamics~\cite{Mielke2014}, 
there are also potential connections of this work to gradient-flow characterisations of Lindblad dynamics~\cite{Mittnenzweig2017}.

\begin{acknowledgements}
FC acknowledges support from a Teach@T\"ubingen Fellowship, through the Deutsche Forschungsgemeinsschaft (DFG, German Research Foundation) under Project No. 435696605, and through the ``Wissenschaftler R\"uckkehrprogramm GSO/CZS” of the Carl-Zeiss-Stiftung and the German Scholars Organization e.V.
JPG acknowledges financial support from EPSRC Grant no.~EP/R04421X/1. JPG is grateful to All Souls College, Oxford, for support through a Visiting Fellowship during the latter stages of this work.
\end{acknowledgements}

\begin{appendix}

\section{Integration and differentiation in $\cal M$}
\label{app:calculus}

This section justifies the definitions in Sec.~\ref{sec:notation}.  The main requirement is to define a suitable integral over pure states.
This is achieved by considering the space $\cal M$ of all $n\times n$ Hermitian matrices so that probability measures (etc) can be defined on this space.
(In practice, all probability measures that we consider are supported on the smaller space ${\cal M}_{\rm p}$, but this does not affect the general theory.)

In this section $\psi$ indicates a generic member of ${\cal M}$ (that is, a Hermitian matrix but not necessarily a density matrix).
It is specified in terms of $n$ real variables (diagonal elements) and $n(n-1)/2$ complex variables (off-diagonal elements).  
To integrate over $\cal M$ we must therefore integrate the $n$ variables over the real line and the complex variables over the whole complex plane.
This suggests that we should define the integration measure $d\psi$ as
\beq
\int d\psi f(\psi) = \int \Big[\prod_{i} d\psi_{ii}\Big] \Big[\prod_{i<j} d(\psi_{ij},\psi_{ij}^*)\Big] f(\psi) \; .
\eeq
where $d(z,z^*)$ indicates that the complex number $z$ is integrated over the whole complex plane.
However, since $\psi_{ij}^*=\psi_{ji}$, we can write this (at least formally) as
\beq
\int d\psi f(\psi) = \int \Big[\prod_{ij} d\psi_{ij}\Big] f(\psi) \; .
\label{equ:meas}
\eeq
That is, instead of dealing explicitly with variables and their complex conjugates (as would usually be required in complex integration), we simply treat each element as an independent variable.  

Using the integration measure (\ref{equ:meas}), it is natural to define (formally) $\delta(\psi-\chi) = \prod_{ij} \delta(\psi_{ij}-\chi_{ij})$, bearing in mind that 
$\delta(\psi_{ij}-\chi_{ij})\delta(\psi_{ji}-\chi_{ji})$ is to be interpreted as $\delta(\psi_{ij}-\chi_{ij})\delta(\psi_{ij}^*-\chi_{ij}^*)$, which satisfies the standard equality
\beq
\int d(\psi_{ij},\psi_{ij}^*) f(\psi_{ij},\psi_{ij}^*) \delta(\psi_{ij}-\chi_{ij})\delta(\psi_{ij}^*-\chi_{ij}^*) =f(\chi_{ij},\chi_{ij}^*)  \;.
\eeq
Hence (\ref{equ:delta}) follows.

A similar situation holds for differentiation.  In order to consider real-valued functions with complex arguments one should generically write $f=f(z,z^*)$ and consider derivatives with respect to both $z$ and $z^*$.  For Hermitian matrices the natural first-order Taylor expansion (separating explicitly the real and complex variables) would be
\beq
f(\psi + \delta\psi) = f(\psi) + \sum_i \delta\psi_{ii} \frac{\partial f}{\partial \psi_{ii} } + 
\sum_{i<j} \left[ \delta\psi_{ij} \frac{\partial f}{\partial \psi_{ij}}  + \delta\psi_{ij}^* \frac{\partial f}{\partial \psi_{ij}^*} \right] + O(\delta\psi)^2 \; .
\eeq
However, using again that $\psi$ and $\delta\psi$ are both Hermitian, we write this as
\beq
f(\psi + \delta\psi) = f(\psi) + \sum_{ij} \delta\psi_{ij} \frac{\partial f}{\partial \psi_{ij} } + O(\delta\psi)^2 \; 
\eeq
which we can abbreviate using (\ref{equ:dot}) as
\beq
f(\psi + \delta\psi) = f(\psi) + \delta\psi \cdot \nabla f  + O(\delta\psi)^2 \; .
\eeq
Given these definitions, one may verify the integration-by-parts formula (\ref{equ:parts}) by using the standard result
\beq
\int d(z,z^*) \left[  f \frac{\partial G}{\partial z}+ f^* \frac{\partial G}{\partial z^*}  \right]
= - \int d(z,z^*) G \left[  \frac{\partial f}{\partial z} +\frac{\partial f^*}{\partial z^*} \right]
\; .
\eeq

\section{level 2.5 for quantum jumps}
\label{app:jump-LD}

\subsection{Tilted generator}
\label{app:jump-tilt}

We show how the SCGF for $(\mu,k)$ in quantum jump processes can be connected to the largest eigenvalue of a tilted generator.
Given a trajectory and two functions $u_1,u_2$, we consider the time-dependence of functions of the form 
\beq
f(\psi_\tau,\mu_\tau,k_\tau) = h(\psi_\tau) \exp\left( -\tau \int d\psi \mu_\tau(\psi) u_1(\psi) - \tau \sum_i \int d\psi d\psi' k^i_\tau(\psi,\psi') u_2^i(\psi,\psi') \right)
\eeq
The increment of this function in a short time interval $[\tau,\tau+dt]$ is 
\begin{multline}
df =  ({\cal B}\cdot \nabla f)  dt 
- u_1(\psi_\tau) f(\psi_\tau,\mu_\tau,k_\tau) dt
\\
+ \sum_i
 [ e^{-u_2^i\left(\psi_\tau,\psi_{+}^i \right)} f(\psi_{+}^i,\mu_\tau,k_\tau) - f(\psi_{\tau},\mu_\tau,k_\tau)] \mathrm{d}n_{i\tau} 
\end{multline}
where $\psi_+^i$ is the jump destination given in (\ref{equ:jump-dest}).
The expectation of $df$ is
\begin{multline}
\mathbb{E}[df] = \mathbb{E}\Big[ {\cal B}\cdot \nabla f - u_1(\psi_\tau) f(\psi_\tau,\mu_\tau,k_\tau) 
\\ + \int d\psi' \, w_i(\psi,\psi') [ e^{-u_2\left(\psi_\tau,\psi' \right)}  f(\psi',\mu_\tau,k_\tau) - f(\psi_{\tau},\mu_\tau,k_\tau)]\Big] dt
\label{equ:inc-jump}
\end{multline}
Define 
\beq
Z_u(\mu_\tau,k_\tau)=\exp\left( -\tau \int d\psi \mu_\tau(\psi) u_1(\psi) - \tau \sum_i \int d\psi d\psi' k^i_\tau(\psi,\psi') u_2^i(\psi,\psi') \right).
\eeq 
Hence we obtain from (\ref{equ:inc-jump}) that
\beq
\frac{\partial}{\partial\tau} \mathbb{E}[ h(\psi_\tau) Z_u(\mu_\tau,k_\tau) ] = \mathbb{E}\left[ {\cal W}_u[h(\psi_\tau)] Z_u(\mu_\tau,k_\tau) \right] 
\label{equ:gen-h-Z}
\eeq
where ${\cal W}_u$ is the tilted generator (\ref{equ:Wuf}).

Now identify $P_{u,\tau}(\psi) = \mathbb{E}[ \delta(\psi_\tau-\psi) Z_u(\mu_\tau,k_\tau) ]$ as a reweighted (non-normalised) density for $\psi$.
Setting $h(\psi_\tau)=\delta(\psi_\tau-\psi)$ in (\ref{equ:gen-h-Z}) yields
\beq
\frac{\partial}{\partial\tau} P_{u,\tau}(\psi) = {\cal W}^\dag_u[ P_{u,\tau}(\psi) ] 
\label{equ:P-lin-jump}
\eeq
where $ {\cal W}^\dag_u$ is the adjoint of ${\cal W}_u$, specifically
\begin{multline}
{\cal W}^\dag_u[ P(\psi) ] = - \nabla\cdot [ {\cal B}[\psi]  P(\psi) ] - u_1(\psi) P(\psi) 
\\ + \sum_{i}\int d\psi'\, \left[  P(\psi') w_i(\psi',\psi)  e^{-u_2^i(\psi,\psi')} - P(\psi) w_i(\psi,\psi') \right] 
\; .
\label{equ:PsW}
\end{multline}
For large times, the linear differential equation (\ref{equ:P-lin-jump}) is controlled by the largest eigenvalue of ${\cal W}_u^\dag$.
Note also that $G_\tau$ of (\ref{equ:Gfu}) is given by 
\beq
G_\tau[u_1,u_2] = \int d\psi \, P_{u,\tau}(\psi) 
\; .
\eeq
 This provides the expected result: the SCGF $\Theta[u_1,u_2]$ in (\ref{equ:Theta-u}) coincides with the largest eigenvalue of ${\cal W}_u^\dag$ (which is also the largest eigenvalue of ${\cal W}_u$).  

\subsection{Large deviation principle}
\label{app:jump-ratefn}

We solve the supremum in \eqref{equ:I2.5-jump-legendre}.  Denote the quantity to be maximised by
\beq
F[u_1,u_2] = - \Theta[u_1,u_2] -\int d\psi\, u_1(\psi) \mu(\psi) - \sum_i \int d\psi d\psi' u_2^i(\psi,\psi') k^i(\psi,\psi')
\; .
\label{equ:Fuu}
\eeq
Taking functional derivatives, one sees immediately that
\begin{align}
\frac{\delta \Theta}{\delta u_1(\psi)} + \mu(\psi) & = 0
\nonumber \\
\frac{\delta \Theta}{\delta u_2^i(\psi,\psi')} + k^i(\psi,\psi') & = 0
\; .
\label{equ:EL-u}
\end{align}
Now recall that $\Theta$ is the maximal eigenvalue of ${\cal W}_u$; denote the corresponding eigenfunction by $f_R$.   Then
\beq
\Theta f_R(\psi) = {\cal B}[\psi] \cdot \nabla f_R(\psi) - u_1(\psi) f_R(\psi) 
+ \sum_{i}\int d\psi'\,  w_i(\psi,\psi')\left[  e^{-u_2^i(\psi,\psi')} f_R(\psi')-f_R(\psi)\right]
\; .
\label{equ:theta-fR}
\eeq
Similarly let $f_L$ be the corresponding eigenfunction of ${\cal W}_u^\dag$ so that
\begin{multline}
\Theta f_L(\psi) = - \nabla\cdot [ {\cal B}[\psi] f_L(\psi) ] - u_1(\psi) f_L(\psi) 
\\ + \sum_{i}\int d\psi'\, \left[  f_L(\psi') w_i(\psi',\psi)  e^{-u_2^i(\psi,\psi')} - f_L(\psi) w_i(\psi,\psi') \right]
\; .
\label{equ:theta-fL}
\end{multline}
We normalise these eigenfunctions such that $\int d\psi\, f_L(\psi) f_R(\psi)=1$.  

Multiplying (\ref{equ:theta-fR}) by $f_L(\psi)$ and integrating with respect to $\psi$ yields
\beq
\Theta = \int d\psi \, f_L {\cal B} \cdot \nabla f_R - \int d\psi \, u_1 f_L f_R + \int d\psi d\psi' \, f_L(\psi) w(\psi,\psi') \left[ e^{-u_2^i(\psi,\psi')}  f_R(\psi')-f_R(\psi)\right]
\; .
\label{equ:theta-symm}
\eeq
(For compactness of notation, we omit the arguments of some functions, in cases where there is no ambiguity.)
We now take a functional derivative with respect to $u_1$.  Note that the eigenfunctions $f_L,f_R$ depend on $(u_1,u_2)$: the result can be written as
\beq
\frac{\delta\Theta}{\delta u_1(\psi')} + f_L(\psi') f_R(\psi') 
= \int d\psi \left[ {\cal W}_u[f_R(\psi)] \frac{\delta}{\delta u_1(\psi')} f_L(\psi) 
+  {\cal W}_u^\dag[f_L(\psi)] \frac{\delta}{\delta u_1(\psi')} f_R(\psi) \right]
\; .
\eeq
Using that $f_R,f_L$ are eigenfunctions of ${\cal W}_u,{\cal W}_u^\dag$ respectively, the right hand side is \\ $\Theta (\delta/\delta u_1) \int d\psi f_L(\psi)  f_R(\psi)$ which vanishes by normalisation of the eigenfunctions.  Combining with (\ref{equ:EL-u}) gives
\beq
\mu(\psi) = f_L(\psi) f_R(\psi)
\; .
\label{equ:muLR}
\eeq
Similarly, differentiating (\ref{equ:theta-symm}) with respect to $u_2$ one obtains [using (\ref{equ:EL-u})]
\beq
k^i(\psi,\psi') = f_L(\psi) w_i(\psi,\psi') e^{-u_2^i(\psi,\psi')} f_R(\psi') 
\; .
\label{equ:k-aux}
\eeq
This is an important result: it says that the empirical jump rate from $\psi$ to $\psi'$ can be expressed as
$
k^i(\psi,\psi') = \mu(\psi) w^A_i(\psi,\psi')
$
where 
\beq
w^A_i(\psi,\psi') = f_R(\psi)^{-1} w_i(\psi,\psi') e^{-u_2^i(\psi,\psi')} f_R(\psi') 
\label{equ:wA-fR}
\eeq 
is an auxiliary jump rate, see Sec.~\ref{sec:doob-jump-unravel}.

Now multiply (\ref{equ:theta-fR}) by $f_L(\psi)$; also multiply (\ref{equ:theta-fL}) by $f_R(\psi)$; and subtract the results.  We obtain
\begin{multline}
0 = f_L {\cal B} \cdot \nabla f_R + f_R \nabla\cdot [ {\cal B} f_L ] + \sum_{i}\int d\psi'\,  f_L(\psi) w_i(\psi,\psi')  e^{-u_2^i(\psi,\psi')} f_R(\psi')
\\ -  \sum_{i}\int d\psi'\,  f_L(\psi') w_i(\psi',\psi)  e^{-u_2^i(\psi,\psi')} f_R(\psi)
\label{equ:cont-intermediate}
\end{multline}
which reduces [using (\ref{equ:muLR},\ref{equ:k-aux})] to
the continuity condition \eqref{CES}.  It follows that finding a supremum in \eqref{equ:I2.5-jump-legendre} requires that \eqref{CES} holds.  (In other cases the supremum is $+\infty$ so the rate function is infinite.)

Finally, combining (\ref{equ:theta-symm}) with (\ref{equ:muLR},\ref{equ:k-aux}) and (\ref{equ:Fuu}) one obtains
\beq
F = - \int d\psi \, f_L {\cal B} \cdot \nabla f_R 
- \sum_i \int d\psi d\psi' \, \left[ k^i(\psi,\psi') - \mu(\psi) w(\psi,\psi')
-  u_2^i(\psi,\psi') k^i(\psi,\psi') \right] 
\; .
\eeq
Rearranging (\ref{equ:k-aux}) we substitute for $u_2$.  After some manipulations we obtain
\begin{multline}
F = \sum_i \int d\psi d\psi' \, {\rm D}\Big[k^i(\psi,\psi') \Big| \mu(\psi)w_i(\psi,\psi')\Big] 
\\ + \int d\psi \,  ( \nabla \cdot [ \mu {\cal B} ]  )  \log f_R 
- \sum_i \int d\psi d\psi' \,  k^i(\psi,\psi') \log \frac{f_R(\psi)}{f_R(\psi')} 
\; 
\end{multline}
where ${\rm D}$ was defined in (\ref{equ:Dxy}).  Using the continuity condition \eqref{CES} one finds that the terms in the second line cancel each other; hence the maximal value of $F$ is indeed given by (\ref{I25qu}), as required.

\section{Quantum Doob transform}
\label{app:doob-mean}

We show that the unravelled dynamics of the quantum Doob process  \eqref{Av-Doob} coincides with $\Psi_t$ defined in \eqref{equ:Psi}.
It is sufficient to show that the expectation value of $\Psi_t$ follows the Lindblad evolution of the quantum Doob process.
Denote this average (under the auxiliary dynamics) by $\rho_\lambda^D(t)=\mathbb{E}[\Psi_t]$;
its time-dependence can be obtained from the generator ${\cal W}^A_\lambda$.  
We interpret $\Psi$ in \eqref{equ:Psi} as a function, that is $\Psi(\psi) = \ell^{1/2} \psi \ell^{1/2}/\Tr(\ell\psi)$.  
Then (\ref{equ:gen}) yields
\beq
\frac{d}{dt} \rho_\lambda^D(t) = \mathbb{E}\left[ {\cal W}^A_\lambda[\Psi(\psi_t)] \right]
\label{equ:rhoD-WA}
\eeq
and recalling \eqref{equ:W-Doob-lambda} one finds
\beq
{\cal W}^A_\lambda[\Psi(\psi)] =  \frac{{\cal W}_\lambda[ \ell^{1/2} \psi \ell^{1/2} ]}{\Tr(\ell\psi)}  -  \theta_k(\lambda) \Psi(\psi) \; .
\eeq
Using  that the generator is linear and also (\ref{equ:W-lambda-L}) gives
\beq
{\cal W}^A_\lambda[\Psi(\psi)] =  \frac{ \ell^{1/2} {\cal L}_\lambda(\psi) \ell^{1/2} }{ \Tr(\ell\psi) } -  \theta_k(\lambda) \Psi(\psi) \; .
\eeq
Hence by \eqref{equ:gen} and using linearity of ${\cal L}_\lambda$
\beq
\frac{d}{dt} \rho^D_\lambda(t) = \mathbb{E}\left[  \ell^{1/2} {\cal L}_\lambda \left(\frac{ \psi_t}{\Tr(\ell\psi_t) } \right) \ell^{1/2} -  \theta_k(\lambda) \Psi(\psi_t) \right]  \; .
\eeq
Finally, re-expressing $\psi$ in terms of $\Psi$ and taking the expectation:
\beq
\frac{d}{dt} \rho^D_\lambda(t) =  \ell^{1/2} {\cal L}_\lambda\left( \ell^{-1/2} \rho^D_\lambda(t) \ell^{-1/2} \right) \ell^{1/2}  -  \theta_k(\lambda) \rho^D_\lambda(t) 
\eeq
from which we recognise $\frac{d}{dt} \rho^D_\lambda = {\cal L}^D_\lambda( \rho^D_\lambda ) $ as required: the average of $\Psi$ follows the quantum Doob dynamics. \\ 

Recalling Fig.~\ref{fig:doob}, the unravelled master equation associated with the quantum Doob dynamics can be obtained in two ways. One option is to unravel the Lindblad dynamics ${\cal L}^D_\lambda$.  The other is to derive the unravelled generator ${\cal W}^D_\lambda$, by considering the time evolution of $\Psi$.  The remainder of this section uses this latter calculation to confirm that the two options yield consistent results.  

By analogy with (\ref{equ:gen}), the generator ${\cal W}^D_\lambda$ obeys
\beq
\frac{d}{dt}\mathbb{E}[ f(\Psi_t) ] = \mathbb{E}\left[ {\cal W}^D_\lambda[f(\Psi_t)] \right]
 \; .
 \label{equ:genW-doob}
 \eeq  
This allows ${\cal W}^D_\lambda$ to be derived following a similar approach to (\ref{equ:rhoD-WA}), which amounts to a change of variables.
Define a function $g_f(\psi) = f(\Psi(\psi))$, so the left hand side of (\ref{equ:genW-doob}) can be expressed as $\mathbb{E}[{\cal W}^A[g_f(\psi)]]$.
The generator (\ref{equ:W-Doob-lambda}) of the auxiliary process is
\begin{equation}
{\cal W}_\lambda^A[g_f(\psi)]= \mathcal{B}[\psi]\cdot \nabla g_f(\psi)+\sum_i\int d\psi'w_i^A(\psi,\psi')\left[g_f(\psi')-g_f(\psi)\right]\, .
\label{aux-to-D}
\end{equation}

The two terms on the right hand side of (\ref{aux-to-D}) will be considered separately.
Using (\ref{equ:dot}) and the multivariate chain rule, the first term is
\beq
\mathcal{B}[\psi]\cdot \nabla g_f(\psi)=\sum_{ij}\left(\mathcal{B}[\psi]\right)_{ij}\frac{\partial g_f}{\partial \psi_{ij}}=\sum_{ij,hk}\left(\mathcal{B}[\psi]\right)_{ij}\frac{\partial \Psi_{hk}}{\partial \psi_{ij}}\frac{\partial f}{\partial \Psi_{hk}} \, .
\label{equ:app-drift}
\eeq
From (\ref{equ:Psi}) we have 
$$
\frac{\partial \Psi_{hk}}{\partial \psi_{ij}}=\frac{(\ell^{1/2})_{hi}(\ell^{1/2})_{jk}}{\Tr(\ell \psi)}-\frac{\left(\ell^{1/2}\psi\ell^{1/2}\right)_{hk}}{(\Tr(\ell \psi))^2}\ell_{ij}\, , 
$$
which yields
\begin{equation*}
\sum_{ij}(\mathcal{B}[\psi])_{ij}\frac{\partial \Psi_{hk}}{\partial \psi_{ij}}=\frac{(\ell^{1/2}\mathcal{B}[\psi]\ell^{1/2})_{hk}}{\Tr(\ell \psi)}-\frac{(\ell^{1/2}\psi\ell^{1/2})_{hk}}{(\Tr(\ell\psi))^2}\Tr(\ell^{1/2}\mathcal{B}[\psi]\ell^{1/2})\, .
\end{equation*}
Now define 
$
\mathcal{H}[\psi]=-iH_{\rm eff}\psi+i\psi H_{\rm eff}^\dagger
$
and use the definition of ${\cal B}$ from (\ref{equ:cal-B}) to obtain
\begin{equation}
\sum_{ij}(\mathcal{B}[\psi])_{ij}\frac{\partial \Psi_{hk}}{\partial \psi_{ij}}=\left(\frac{\ell^{1/2}\mathcal{H}[\psi]\ell^{1/2}}{\Tr(\ell \psi)}-\frac{\ell^{1/2}\psi\ell^{1/2}}{(\Tr(\ell\psi))^2}\Tr(\ell\mathcal{H}[\psi])\right)_{hk}\, .
\label{drift-W^D}
\end{equation}

Following~\cite{carollo2019}, the next step is to re-express this formula in terms of $\Psi$, to obtain $\mathcal{W}_\lambda^D$.
Define $\tilde{H}_{\rm eff}=\tilde{H}-i/2\sum_i\tilde{L}^\dagger_i \tilde{L}_i$ [similar to \eqref{equ:doob-HJ}]
and 
$$
\tilde{B}[\Psi]=-i\tilde{H}_{\rm eff}\Psi+i\Psi\tilde{H}_{\rm eff}^\dagger-\Psi\Tr(-i\tilde{H}_{\rm eff}\Psi+i\Psi\tilde{H}_{\rm eff}^\dagger)\, .
$$
which is the analogue of (\ref{equ:cal-B}) for the unravelled quantum Doob dynamics.  Then (\ref{drift-W^D}) can be expressed as
\beq
\sum_{ij}(\mathcal{B}[\psi])_{ij}\frac{\partial \Psi_{hk}}{\partial \psi_{ij}}=(\tilde{\mathcal{B}}[\Psi])_{hk}\, ,
\eeq
so (\ref{equ:app-drift}) becomes
\beq
\mathcal{B}[\psi]\cdot \nabla g_f(\psi)=\tilde{\mathcal{B}}[\Psi]\cdot \nabla f(\Psi)\, .
\label{equ:app-drift-Psi}
\eeq

Now consider the second term on the right hand side of \eqref{aux-to-D}: one uses (\ref{equ:wA-ell}), performs the integral over $\psi'$, and re-expresses $\psi$ in terms of $\Psi$, to obtain
\begin{equation}
\sum_i\int d\psi'w_i^A(\psi,\psi')\left[g_f(\psi')-g_f(\psi)\right]
= \sum_i\int d\Psi'\tilde{w}_i(\Psi,\Psi')\left[f(\Psi)-f(\Psi')\right]\, ,
\label{equ:app-jump-Psi}
\end{equation}
with [recalling (\ref{equ:doob-HJ})]
$$
\tilde{w}_i(\Psi,\Psi')=\Tr(\tilde{L}_i\Psi\tilde{L}_i^\dagger )\delta\left(\Psi'-\frac{\tilde{L}_i\Psi \tilde{L}_i^\dagger}{\Tr(\tilde{L}_i\Psi \tilde{L}_i^\dagger)}\right)\, .
$$

Finally using (\ref{equ:app-drift-Psi},\ref{equ:app-jump-Psi}) with (\ref{aux-to-D}) accomplishes the change of variable from $\psi$ to $\Psi$: we have
\beq
{\cal W}_\lambda^A[g_f(\psi)] = \mathcal{W}_\lambda^D[f(\Psi)]
\label{equ:WA-WD}
\eeq
with
\beq
\mathcal{W}_\lambda^D[f(\Psi)]=\tilde{\mathcal{B}}[\Psi]\cdot \nabla f(\Psi)+\sum_i\int d\Psi'\tilde{w}_i(\Psi,\Psi')\left[f(\Psi)-f(\Psi')\right]\, .
\eeq
Taking the expectation of (\ref{equ:WA-WD}) and comparing with (\ref{equ:genW-doob}) confirms that this $\mathcal{W}_\lambda^D$ is the generator for the stochastic process $\Psi_t$.  
 Since the jump operators $\tilde L$ and the Hamiltonian $\tilde H$ are those of the quantum Doob dynamics, this shows that the same process $\Psi$ can be obtained either by unravelling the quantum Doob dynamics, or by constructing the auxiliary process (\ref{equ:wA-k-mu}) and performing the change of variable (\ref{equ:Psi}).
 This establishes the status of $\mathcal{W}_\lambda^D$ as illustrated in Fig.~\ref{fig:doob}.

\section{Quantum reset processes and thermodynamic uncertainty relation}
\label{app:qu-reset}

The level 2.5 formalism can prove very useful when aiming at establishing bounds to large fluctuations of some time-integrated observables. In particular, it can provide thermodynamic uncertainty relations for jump rates~\cite{Gingrich2016}, and for other types of observables which cannot be addressed by means of tilted operator techniques. Here we review the discussion of Ref.~\cite{carollo2019}, explaining how these ideas work for quantum reset process.

A quantum reset process is a Lindblad quantum process where the destination state of
each jump operator $L_i$ in (\ref{equ:jump-dest}) is independent of the initial point $\psi$ of the jump.  That is
\beq
L_i\psi L_i^\dagger = \varphi_i \Tr(L_i\psi L_i^\dagger) \, ,
\eeq
where $\varphi_i$ is a fixed matrix (the destination state).

For such processes, the theory for unravelled processes simplifies considerably.
Every quantum jump \emph{resets} the system to one of the states $\varphi_i$ (for $i=1,2,\dots,M$), we refer to these as \emph{reset states}.
From \eqref{equ:psi-det}, if the last jump occurred at time $t$ and was of type $i$ then the state of the system at time $t+t'$
is 
\beq
\varphi_{i}(t')=\frac{e^{-i H_{\rm eff} t'}\varphi_i e^{i H_{\rm eff}^\dagger t'}}{\Tr\left[e^{-i H_{\rm eff} t'}\varphi_i e^{i H_{\rm eff}^\dagger t'}\right]} \; .
\label{equ:vhi-t}
\eeq
Also, let $S_i(t')$ be the probability that this system survives up to time $t+t'$ without jumping.  This evolves in time as
\beq
\frac{d}{dt'}S_i(t')=-\sum_j p_{ij}(t')
\label{equ:S-p}
\eeq 
where
\beq
p_{ij}(t') dt' = S_i(t') dt' \int d\psi' w_j(\varphi_i(t'),\psi') 
\label{equ:pij-w}
\eeq
is the probability to make a jump of type $j$ in the time interval $[t+t';t+t'+dt']$.  (This is the product of the probability to make no jump in $[t,t+t']$ and the rate to jump at time $t'$.  It is normalised as $\sum_j \int_0^\infty dt\, p_{ij}(t) = 1$.)  

The functions $p_{ij}(t)$ and $\varphi_i(t)$ fully specify the quantum reset process.  The survival probabilities $S_i(t)$ are related to the $p_{ij}$ by (\ref{equ:S-p}).  
We also define the (marginal) probability that the next jump is of type $j$, given that the last one was of type $i$:
\beq
R_{ij}=\int_0^\infty dt'\, p_{ij}(t')\, ,
\label{equ:RR}
\eeq
with $\sum_j R_{ij}=1$.
Also let $c_i=\mathbb{E}[\bar k^i_\tau]=\int d\psi d\psi' \Gamma_i(\psi,\psi')$ be the \emph{total} rate of jumps of type $i$ (independent of the type of the last jump), recall (\ref{equ:Gamma}).  
So
\beq
c_i = \int d\psi d\psi' P_{\infty}(\psi) w_i(\psi,\psi') \; .
\eeq

With these definitions in hand, we now characterise the steady state of the quantum reset process.
Since the evolution between jumps is deterministic as in \eqref{equ:vhi-t}, it follows that the steady state is supported on the deterministic paths that start from each reset state, that is
\beq
P_{\infty}(\psi) = \sum_i c_i \int_0^\infty dt\, S_i(t) \delta\!\left( \psi - \varphi_i(t) \right) \; .
\label{equ:Pinf-reset}
\eeq
The jump rate $c_i$ in the above formula is the rate at which the system jumps via channel $i$, i.e. jumps into $\varphi_i$. With the above $P_\infty$, using the definition of $c_i$, we can compute 
\beq
c_i=\sum_j c_j R_{ji}\, .
\label{equ:c-R-eigen}
\eeq
This is an eigenvalue problem for the matrix $R$. The normalisation $\sum_j R_{ij}=1$ and the fact that all $R_{ij}$ are non-negative mean that there is a solution with $c_i>0$ for all $i$, as required.  This fixes the $c_i$ up to an overall multiplicative constant which can be found by insisting that $P_\infty$ is normalised in \eqref{equ:Pinf-reset}.
It may be verified that $P_\infty$  is indeed the steady state of \eqref{UQME}. 

Physically, the statistical weight of $\varphi_i(t)$ (in the steady state) is the product of the rate of jumps into $\varphi_i$ and the probability to survive a time $t$ before jumping again.  


Following~\cite{carollo2019}, we now derive a thermodynamic uncertainty relation (TUR) for quantum reset processes. 
This concerns full-counting statistics at level-$1$.
(For other quantum TURs see also Refs.~\cite{PhysRevLett.120.090601,PhysRevResearch.1.033021,PhysRevLett.125.050601,PhysRevB.101.195423}.)
Combining (\ref{equ:I1-contract},\ref{equ:I2.5-wA}) one sees that
\beq
I_1(\bar k) = \inf_{w^A} \left( \sum_i\int d\psi d\psi' \mu(\psi){\rm D}\Big[w_i^A(\psi,\psi')\Big|w_i(\psi,\psi')\Big] \right)
\label{equ:I1-contract-wA}
\eeq
where the infimum is constrained such that the auxiliary process (with rates $w^A$) must have an average jump rate $\bar k$, and $\mu$ is the steady state distribution of the auxiliary process.
As explained in Sec.~\ref{sec:quantum-doob}, the solution to this infimum is obtained when $w^A$ is given by \eqref{equ:wA-ell}, but we do not have explicit formulae for the quantities appearing in that equation.  However, bounds on $I_1$ are available from \eqref{equ:I1-contract-wA}, by choosing suitable auxiliary processes that solve the constraint.

The auxiliary process that we consider is also a quantum reset process. 
 It has the same reset states and the same deterministic evolution, so the function $\varphi_i(t)$ remains the same.  The analogue of the function $p_{ij}(t)$ is $\hat{p}_{ij}(t)$.  This fully determines the auxiliary quantum reset process and we define the corresponding $\hat{S}_i(t)$, $\hat{R}_{ij}(t)$ and $\hat{c}_i$ by analogy with $S_i,R_{ij},c_i$ from above.

From \eqref{equ:pij-w}, one sees that the corresponding jump rate is
\beq
w_j^A(\varphi_i(t) ,\psi')= \frac{ \hat{p}_{ij}(t) }{ \hat{S}_i(t) } \delta( \psi' - \varphi_j )
\eeq
(It is sufficient to specify this rate for states $\psi$ within the support of $P_\infty$ so this fully determines the auxiliary jump rates.)  
The steady state distribution of the auxiliary process is $\mu$, which is given by \eqref{equ:Pinf-reset}, with $(c,S)$ replaced by $(\hat{c},\hat{S})$.
Hence for this process one has from (\ref{equ:I2.5-wA}) that 
\beq
I_{2.5}(\mu,k) = \sum_{ij} \hat{c}_i  \int_0^\infty dt \, \left[ \hat{p}_{ij} \log \frac{ \hat{p}_{ij} }{ p_{ij} } - \hat{p}_{ij}  \log \frac{ \hat{S}_{i} }{ S_{i} }
 -  \hat{p}_{ij} +  \hat{S}_{i} \frac{p_{ij} }{ S_{i} } \right] \, .
\eeq
Using (\ref{equ:S-p}) one has $\sum_j {p}_{ij} = -(d{S}_i/dt) $ and similarly for $\hat{p},\hat{S}$.  This allows the sum over $j$ to be performed in some terms, yielding
\begin{multline}
I_{2.5}(\mu,k) = \sum_{ij} \hat{c}_i  \int_0^\infty dt \,  \hat{p}_{ij} \log \frac{ \hat{p}_{ij} }{ p_{ij} } 
+\sum_i \hat{c}_i \int_0^\infty dt \left[  \frac{d\hat{S}_i}{dt}  \Big( \log \frac{ \hat{S}_{i} }{ S_{i} } + 1 \Big)
 -  \hat{S}_{i} \frac{d}{dt} \log{S}_i  \right] 
\end{multline}
After some integrations by parts and using $S_i(0)=1$ and $S_i(\infty)=0$, the second term on the right-hand side evaluates to zero.
Moreover, this auxiliary model has average jump rates $\bar{k}^i = \hat{c}_i$.  Hence by
(\ref{equ:I1-contract-wA}), we find
\beq
I_1(\bar{k})\le\sum_{i,j}\bar{k}^i\int_0^\infty dt\, \hat{p}_{ij}(t)\log \frac{\hat{p}_{ij}(t)}{p_{ij}(t)}\, .
\label{equ:I1-bound-pij}
\eeq
This is a generic bound on the rate function at level-1, as long as the $\hat{p}_{ij}$ are chosen such that the resulting $\hat{c}_i=\bar{k}^i$.

We now make a specific choice for $\hat{p}$, as
\beq
\hat{p}_{ij}(t)=v_{ij}e^{-u_{ij} t}p_{ij}(t)\,,
\label{equ:p-uv}
\eeq
where $u_{ij},v_{ij}$ are chosen to satisfy $\hat{R}_{ij}=R_{ij}$, as we now discuss.  (The idea to accelerate or reduce the jump rate, leaving the distribution of jump destinations invariant.)  
To find the relation between $u$ and $v$, recall (\ref{equ:RR}) and define
\beq
\tau_{ij}=\frac{\int dt \, t \, p_{ij}(t)}{R_{ij}}\, , \qquad  \mbox{ and } \qquad \sigma^2_{ij}=\frac{\int dt (t^2-\tau_{ij}^2)  p_{ij}(t)}{R_{ij}} \, .
\eeq
This $\tau_{ij}$ is the mean time between a jump of type $i$ and one of type $j$, and $\sigma_{ij}^2$ is the variance of this time.
Then by (\ref{equ:p-uv}) one
has
\beq
\hat{R}_{ij} = v_{ij} R_{ij} \left[ 1 - u_{ij} \tau_{ij} + (u_{ij}^2/2) ( \sigma_{ij}^2 + \tau_{ij}^2 ) + O(u_{ij}^3) \right]
\eeq
We take $v_{ij} = 1 + u_{ij} \tau_{ij} + u_{ij}^2 (\tau_{ij}^2-\sigma_{ij}^2)/2$ which ensures $\hat{R}_{ij} = R_{ij}(1+O(u^3))$.  
By the eigenproblems \eqref{equ:c-R-eigen} for $c,\hat{c}$, this means that  $\hat{c}_i=c_i(\lambda+O(u^3))$ for some constant $\lambda$ (independent of $i$).  This $\lambda$ is the factor by which the auxiliary dynamics has been accelerated.  

The conditions $\hat{R}_{ij}=R_{ij}$ from above have been used to fix the $v_{ij}$, 
but the $u_{ij}$ are still free.  We choose these to achieve a uniform acceleration of all jumps, as usual in derivations of TURs.  To achieve this define $\hat\tau_{ij}$ analogous to $\tau_{ij}$ and choose $u_{ij}$ such that $\hat\tau_{ij}=\tau_{ij}/\lambda$.  This requires $u_{ij} = (\lambda-1)\tau_{ij}/\sigma_{ij}^2$.  Now $\hat{p}_{ij}$ is fully determined in terms of $\lambda$ and the auxiliary process has average jump rates $\bar{k}^i = \lambda c_i$.  It follows that $ \int \hat p_{ij} \log (\hat p_{ij} / p_{ij}) dt = R_{ij} \tau_{ij}^2 ( \lambda-1)^2 / (2\sigma_{ij}^2)$, with a correction of order $O(\lambda-1)^3$.  Using this in (\ref{equ:I1-bound-pij}) yields
\beq
I_1(\lambda c) \leq \frac{\chi}{2}(\lambda-1)^2  +O (\lambda-1 )^3 \, ,
\label{equ:I1-lambda-bound}
\eeq
where $c$ is a vector whose elements are the $c_i$, and
\beq
 \chi=\sum_{ij} c_i R_{ij}\frac{\tau_{ij}^2}{\sigma_{ij}^2}\, .
\eeq
Note that $I_1$ in \eqref{equ:I1-lambda-bound} is the rate function from \eqref{equ:I1-contract-wA}, whose argument is the vector $\bar k$.  The inequality \eqref{equ:I1-lambda-bound} applies when the argument $\bar k$ of the rate function is parallel to $c$.

Now consider an arbitrary linear combination of jump rates $k^b_\tau = \sum b_i\bar{k}^i_\tau$ and let $b$ be the vector whose elements are the $b_i$.
Clearly $\mathbb{E}[k^b_\tau]=b\cdot c$.  For large $\tau$ then $k^b_\tau$ obeys a central limit theorem whose variance can be obtained from the rate function $I_1$, specifically, $\tau\mathrm{Var}(k^b_\tau ) \approx b \cdot {\cal H}^{-1} b$ where ${\cal H}$ is the Hessian of $I_1$ (at its minimum).  Since ${\cal H}$ is symmetric positive definite then the Cauchy-Schwartz inequality implies that $( b \cdot {\cal H}^{-1} b) ( c \cdot {\cal H } c ) \geq ( b\cdot c)^2$.  From \eqref{equ:I1-lambda-bound} then $( c \cdot {\cal H } c ) \le  \chi$ so
one has a thermodynamic uncertainty relation
\beq
\lim_{\tau\to\infty} \frac{\tau\,\mathrm{Var}(k^b_\tau ) }{ \mathbb{E}(k^b_\tau)^2 } \geq \frac{1}{\chi} \; .
\label{equ:jump-TUR}
\eeq
For classical systems $p_{ij}$ is an exponential distribution so $\sigma_{ij}=\tau_{ij}$ and $\chi = \sum_i \bar{k}^i$ is the total jump rate, so (\ref{equ:jump-TUR}) becomes a classical TUR~\cite{Garrahan2017}.
For quantum systems one may have anti-bunching of jump events, leading to $\sigma_{ij}<\tau_{ij}$.  In such cases the variance of time-averaged currents can violate the classical thermodynamic uncertainty relation while still obeying \eqref{equ:jump-TUR}.  This theory was analysed for a simple model system in~\cite{carollo2019}.

\section{Quantum diffusions}

\subsection{Tilted generator}
\label{app:diff-tilt}

We derive the tilted generator (\ref{equ:Theta-a}) whose largest eigenvalue is $\Theta[a_1,a_2]$ in (\ref{equ:Theta-a}).
The structure of the calculation is the same as that of Appendix~\ref{app:jump-tilt}.
Given a trajectory and two functions $a_1,a_2$, we consider the time-dependence of functions of the form 
\beq
f(\psi_\tau,\mu_\tau,j_\tau) = h(\psi_\tau) \exp\left( \tau \int d\psi \mu_\tau(\psi) a_1(\psi) + \tau \sum_m \int d\psi j^m_\tau(\psi) a_2^m(\psi) \right)
\; .
\eeq
The increment of this function in a short time interval $[\tau,\tau+dt]$ is [similar to (\ref{Stoc-Fun})]
\begin{multline}
df = \sum_{ij}\frac{\partial f}{\partial \psi_{ij}}  (d \psi_t)_{ij}
+\frac{1}{2}\sum_{ij,hk}\frac{\partial^2 f}{\partial \psi_{ij} \partial \psi_{hk}}(d \psi_t)_{ij}(d \psi_t)_{hk}
+a_1(\psi) f dt
+ \sum_m a_2^m(\psi) f dW^m_t 
\\
+ \frac12 \sum_{mn} a_2^m(\psi) a_2^n(\psi) f dW^m_t dW^n_t
+ \sum_{m,ij} a_2^m(\psi) \frac{\partial f}{\partial \psi_{ij}}  dW^m_t (d\psi_t)_{ij}
\; .
\end{multline}
Taking the expectation yields
\begin{multline}
\mathbb{E}[df] = 
\mathbb{E}\left[\sum_{ij}\frac{\partial f}{\partial \psi_{ij}} (\mathcal{L}[\psi])_{ij}
+\frac{1}{2}\sum_{ij,hk}\frac{\partial^2 f}{\partial \psi_{ij}\psi_{hk}}D_{ij,hk}(\psi) 
+ a_1(\psi) f\right]
\\
+ \mathbb{E}\left[\frac12 \sum_m a^m_2(\psi)^2 f
+ \sum_m a_2^m(\psi) {\cal K}^m \cdot \nabla f \right]
\; .
\end{multline}
Define
\beq
 Z_a(\mu_\tau,j_\tau)  = \exp\left( \tau \int d\psi \mu_\tau(\psi) a_1(\psi) + \tau \sum_m \int d\psi j^m_\tau(\psi) a_2^m(\psi) \right)
 \; .
\eeq
Hence we obtain
\beq
\frac{\partial}{\partial\tau} \mathbb{E}[ h(\psi_\tau) Z_a(\mu_\tau,j_\tau) ] = \mathbb{E}\left[ {\cal W}_a[h(\psi_\tau)] Z_a(\mu_\tau,j_\tau) \right] 
\; .
\label{equ:gen-h-Z-diff}
\eeq
where ${\cal W}_a$ is the tilted generator (\ref{Tilted-A}).  The argument that the largest eigenvalue of this operator coincides with $\Theta[a_1,a_2]$ then follows exactly as in Appendix~\ref{app:jump-tilt}.

\subsection{Rate function at level 2.5}
\label{app:diff-ratefn}

We derive the rate function at level 2.5 for quantum diffusions, using \eqref{equ:diff-2.5-sup}.
The structure of the calculation is similar to that of Appendix~\ref{app:jump-ratefn}.
The object to be maximised is
\beq
F[a_1,a_2] = - \Theta[a_1,a_2] + \int d\psi\, a_1(\psi) \mu(\psi) + \sum_m \int d\psi a_2^m(\psi) j^m(\psi) \; .
\label{equ:Faa}
\eeq
Taking functional derivatives we have
\begin{align}
\frac{\delta \Theta}{\delta a_1(\psi)} & =  \mu(\psi) \, ,
\nonumber \\
\frac{\delta \Theta}{\delta a_2^m(\psi)}  & = j^m(\psi)\, .
\label{equ:EL-a}
\end{align}
The eigenvalue equation for $\Theta$ is
\beq
\Theta f_R = \left[ \mathcal{L} +\sum_{m}a_2^m \mathcal{K}^m \right] \cdot \nabla f_R
+\frac{1}{2}\sum_{ij,hk}D_{ij,hk}\frac{\partial^2 f_R}{\partial \psi_{ij} \partial \psi_{hk}}
+ a_1 f_R +\frac{1}{2}\sum_m (a_2^m)^2 f_R\, ,
\label{theta-eig-app}
\eeq
and corresponding eigenfunction of the adjoint operator is $f_L$ and $\int d\psi f_L f_R=1$.  Hence
\beq
\Theta = \int d\psi f_L \Big\{ \Big( \mathcal{L} +\sum_{m}a_2^m \mathcal{K}^m \Big) \cdot \nabla f_R
+\frac{1}{2}\sum_{ij,hk}D_{ij,hk}\frac{\partial^2 f_R}{\partial \psi_{ij} \partial \psi_{hk}}
+ a_1 f_R +\frac{1}{2}\sum_m (a_2^m)^2 f_R \Big\} \; .
\label{equ:theta-symm-a}
\eeq
Differentiating with respect to $a_1,a_2$ and using \eqref{equ:EL-a} yields (as before) that
\beq
\mu(\psi) = f_L(\psi) f_R(\psi)
\label{equ:mu-lr-diff}
\eeq
 and also
\beq
j^m = f_L {\cal K}^m \cdot \nabla f_R + a_2^m f_L f_R \; .
\label{equ:jm-am}
\eeq
The continuity condition (\ref{equ:CES-diff}) is obtained similarly to (\ref{equ:cont-intermediate}): multiply (\ref{theta-eig-app}) by $f_L$ and subtract a similar expression obtained via the adjoint equation.  One obtains
\begin{multline}
0 =  f_L \Big[ \mathcal{L} +\sum_{m}a_2^m \mathcal{K}^m \Big] \cdot \nabla f_R 
+ \frac12 f_L D_{ij,hk} \frac{\partial^2}{\partial \psi_{ij}\partial\psi_{hk}}  f_R
\\
+ f_R \nabla\cdot \Big[ \mathcal{L} +\sum_{m}a_2^m \mathcal{K}^m \Big] f_L
- \frac12 f_R \frac{\partial^2}{\partial \psi_{ij}\partial\psi_{hk}} (f_L D_{ij,hk}  ) \; .
\end{multline}
This may be simplified [using (\ref{equ:mu-lr-diff})] as
\beq
0 = \nabla\cdot\Big[ \mu {\cal L} + \mu \sum_m a_2^m {\cal K}^m \Big] 
+ \frac12 \sum_{ij,hk} \frac{\partial}{\partial \psi_{ij}} \left[ f_L D_{ij,hk} \frac{\partial f_R}{\partial \psi_{hk}}   - f_R \frac{\partial }{\partial \psi_{hk}}( f_L D_{ij,hk}) \right] \; .
\eeq
Substituting for $a_2^m$ using (\ref{equ:jm-am}) and using the definition of $D_{ij,hk}$ one obtains the continuity condition \eqref{equ:CES-diff}.

It only remains to combine the ingredients, so as to compute the maximal value of $F[a_1,a_2]$.  Using (\ref{equ:theta-symm-a},\ref{equ:mu-lr-diff}) one obtains
\beq
F=
 \int d\psi \,  \Big[ \sum_m a_2^m j^m - f_L \Big( \mathcal{L} +\sum_{m}a_2^m \mathcal{K}^m \Big) \cdot \nabla f_R
-\frac{1}{2}\sum_{ij,hk} f_L D_{ij,hk}\frac{\partial^2 f_R}{\partial \psi_{ij} \partial \psi_{hk}} 
-\frac{1}{2}\sum_m (a_2^m)^2 \mu   \Big] \; .
\eeq
Note from \eqref{equ:jm-am} that $f_L{\cal K}^m \cdot \nabla f_R=j^m - \mu a_2^m$.  Using this to substitute the term involving ${\cal K}^m$ simplifies this equation:
\beq
F=
 \int d\psi \,  \Big[   - f_L  \mathcal{L} \cdot \nabla f_R
-\frac{1}{2}\sum_{ij,hk} f_L D_{ij,hk}\frac{\partial^2 f_R}{\partial \psi_{ij} \partial \psi_{hk}} + \frac{1}{2}\sum_m (a_2^m)^2 \mu  \Big] \; .
\eeq
Let $f_R=e^\Lambda$ which can be used with (\ref{equ:mu-lr-diff}) to eliminate $f_L,f_R$ in favour of $\mu,\Lambda$.  
We obtain
\beq
F=
 \int d\psi \,  \Big[  - \mu \mathcal{L} \cdot \nabla\Lambda 
-\frac{1}{2}\sum_{ij,hk} \mu D_{ij,hk} \left( \frac{\partial^2 \Lambda}{\partial \psi_{ij} \partial \psi_{hk}}  +  \frac{\partial\Lambda}{\partial \psi_{ij}}  \frac{\partial\Lambda}{\partial \psi_{hk}}\right)
+\frac{1}{2}\sum_m (a_2^m)^2 \mu   \Big] \; .
\eeq
Integrating several times by parts and rearranging terms, we obtain
\beq
F=
 \int d\psi \,  \Big\{ \Lambda \sum_{ij} \frac{\partial}{\partial \psi_{ij}} \left[ \mu \mathcal{L}_{ij}
-\frac{1}{2}\sum_{hk} \frac{\partial}{\partial\psi_{hk}} (\mu D_{ij,hk})   \right] + \frac12 \mu \sum_m \left[  
(a_2^m)^2 -( {\cal K}^m \cdot\nabla\Lambda)^2 \right]  \Big\} \; .
\eeq
where we also used the definition of $D$ in  terms of ${\cal K}$.   The first term may be simplified using the continuity equation (\ref{equ:CES-diff}) and
after one more integration by parts we obtain
\beq
F=
 \int d\psi \,  \Big\{  j^m ({\cal K}^m\cdot\nabla\Lambda) + \frac12 \mu \sum_m \left[ (a_2^m)^2  - ( {\cal K}^m \cdot\nabla\Lambda)^2 
\right]  \Big\} \; .
\eeq
Finally using from \eqref{equ:jm-am} that $a_2^m = (j^m/\mu) - ({\cal K}^m \cdot \nabla \Lambda)$ yields $F=\frac12 \int d\psi \sum_m (j^m)^2/\mu$, as required for (\ref{Q-funct}).

\subsection{The empirical probability current as a function of empirical measure and empirical noise}
\label{app:Je-strato}

We derive (\ref{equ:Je-j}) which relates the empirical current to the empirical noise.
It is convenient to introduce a function $x\colon{\cal M}\to\mathbb{R}$ and define a random (trajectory-dependent) variable
\beq
X_\tau=\int d\psi \, x(\psi){J}_\tau^{\rm e}(\psi)\, .
\label{equ:X-x}
\eeq
Note $X_\tau$ is a matrix-valued quantity, as is $J^e_\tau$.
From (\ref{equ:Je-psi}) we have $X_\tau=\frac{1}{\tau}\int_0^\tau x(\psi_t)\circ d\psi_t$. 
Converting from Stratonovich to Ito integral we obtain an equation for the matrix elements of $X_\tau$:
\beq
[X_\tau]_{ij} = \frac{1}{\tau}\int_0^\tau x(\psi_t) (d\psi_{t})_{ij} + \frac{1}{2\tau} \int_0^\tau \sum_{hk}\frac{\partial x}{\partial \psi_{hk}}D_{ij,hk}(\psi_t) dt
\; .
\eeq
Using (\ref{QSSE}) to substitute for $d\psi_t$ yields
\begin{multline}
[X_\tau]_{ij} = \frac{1}{\tau}\int_0^\tau  \Big\{ x(\psi_t) [{\cal L}(\psi_t)]_{ij} +  \frac12 \sum_{hk}\frac{\partial x}{\partial \psi_{hk}}D_{ij,hk}(\psi_t) \Big\} dt 
\\ + \frac{1}{\tau} \int_0^\tau \sum_m x(\psi_t) [{\cal K}^m(\psi_t)]_{ij} dW^m_t
\; .
\end{multline}
Using the definitions of $\mu_\tau$ and $j_\tau$, this becomes
\begin{multline}
[X_\tau]_{ij}  = \int d\psi \Big\{   x(\psi)  [{\cal L}(\psi)]_{ij} \mu_\tau(\psi) + \frac12 \sum_{hk}\frac{\partial x}{\partial \psi_{hk}}D_{ij,hk}(\psi)\mu_\tau(\psi)  
\\ + \sum_m x(\psi)  [{\cal K}^m(\psi)]_{ij} j^m_\tau(\psi) \Big\}
\; .
\end{multline}
Finally one has from \eqref{equ:X-x} that 
$[{J}^{\rm e}_\tau(\psi)]_{ij}=\frac{\delta}{\delta x(\psi)} [X_\tau]_{ij}$, which yields (\ref{equ:Je-j}), as required.

\end{appendix}

\bibliographystyle{spphys}      
\bibliography{QL25QSD}                

\end{document}